\documentclass[journal]{IEEEtran}

\usepackage{graphicx}
\usepackage{amsmath}
\usepackage{amssymb}
\usepackage[caption=false]{subfig}
\usepackage[noadjust]{cite}
\usepackage{float}
\usepackage{algorithm}
\usepackage{bibentry}
\usepackage{balance}
\usepackage{algorithm}
\usepackage{algorithmicx}
\usepackage{algpseudocode}
\usepackage{xcolor}
\usepackage{balance}
\usepackage{multirow}

\usepackage{amsmath,algorithm,tabularx}

\makeatletter
\newcommand{\multiline}[1]{%
  \begin{tabularx}{\dimexpr\linewidth-\ALG@thistlm}[t]{@{}X@{}}
    #1
  \end{tabularx}
}

\DeclareMathOperator{\Tr}{Tr}
\newtheorem{theorem}{Theorem}

\newtheorem{remark}{Remark}

\begin{document}

\title{Max-Min Fair
Precoder Design and Power Allocation for MU-MIMO NOMA}

\author{Ahmet Zahid Yalcin, Mustafa Kagan Cetin and
        Melda Yuksel~\IEEEmembership{IEEE, Senior Member}
 \thanks{A. Z. Yalcin, M. K. Cetin and M. Yuksel are with TOBB University of Economics and Technology, Ankara 06560, Turkey (email: \{azyalcin, mustafakagancetin, yuksel\}@etu.edu.tr).}       
\thanks{This work has been partially supported by HAVELSAN Inc. under grant HVL-SOZ-19/007.}}

\maketitle

\begin{abstract}
In this paper, a downlink multiple input multiple output (MIMO) non-orthogonal multiple access (NOMA) wireless communication system is considered. In NOMA systems, the base station has unicast data for all users, and multiple users in a group share the same resources. 
The objective is to design transmit precoders and power allocation coefficients jointly that provide max-min fairness (MMF) among the strongest users in each group, while maintaining minimum target rates for all the other users. The problem is solved via two main iterative approaches. The first method is based on semi-definite relaxation (SDR) and successive convex approximation (SCA), and the second method is based on the equivalency between achievable rate and minimum mean square error (MMSE) expressions. For the latter approach, Karush-Kuhn-Tucker (KKT) optimality conditions are derived and the expressions satisfied by the optimal receivers, MMSE weights and the optimal precoders are obtained. Proposed algorithms are compared with rate-splitting (RS), orthogonal multiple access (OMA) and multi-user linear precoding (MULP) schemes in terms of MMF rates, energy efficiency and complexity. It is shown that while RS has the best MMF rates and energy efficiency, the MMSE approach based on KKT optimality conditions has the least complexity. Moreover, the SDR/SCA approach offers an excellent tradeoff. It offers high MMF rates, low complexity and superior energy efficiency.

\end{abstract}



\IEEEpeerreviewmaketitle

\section{Introduction}

The demand for data traffic is steadily increasing and wireless networks of the next decade have to meet the high data rate requirements for many different applications \cite{Andrews2014_WhatWill5GBe}. To handle this high data rate, non-orthogonal multiple access (NOMA) is considered as a breakthrough technique, which enables simultaneous multiple access in the power domain for $5$G wireless networks \cite{Dai2015_NOMAfor5G}. Specifically, downlink NOMA is an application of broadcast channels \cite{Cover1972_BroadcastChannel} and it relies on superposition coding (SPC) at the transmitter to transfer multiple data streams in the same resource block, and successive interference cancellation (SIC) at the receiver to cancel co-channel interference. NOMA has the potential to deliver higher system throughput \cite{Saito2013_NOMAForCellular,Benjebbovu2013_SystemLevelPerformance} and higher ergodic sum capacity \cite{Zheng2017_CapacityComparisonBetweenMIMONOMA}, and to achieve better outage performance \cite{Ding2014_OnThePerformanceOfNOMA} compared to the existing orthogonal multiple access (OMA) techniques. In practical power domain NOMA schemes, more power is allocated to users with poor channel conditions to guarantee their required minimum rates \cite{Yang2016_GeneralPowerAllo}. This way NOMA presents an advantage in providing higher spectral efficiency and fairness. 

\vspace{-0.3cm}
\subsection{Related Work}

In NOMA systems, each user can have a dedicated precoding vector, or a cluster of users can share the same precoding vector. The former has the advantage of custom precoding for each user, but suffers from the rank constraints in the downlink multiuser MIMO broadcast channel \cite{Hanif2016_minMaxMethodForSumRateNOMA}. The latter is not limited by rank, but messages are not individually precoded, so channel gain vectors and precoders are mismatched.

Assuming the transmit signals of each user are coded by a dedicated precoding vector, sum rate maximization, total power minimization and max-min fairness for NOMA systems are studied under different constraints and with different methods in the literature. The paper \cite{Hanif2016_minMaxMethodForSumRateNOMA} solves the sum rate maximization problem by approximating the problem with a minorization-maximization algorithm. The paper \cite{Zhu2018_BeamformingDesignforDownlinkNOMA} presents a precoding design for maximizing the sum rate of all users under decoding order and quality-of-service (QoS) constraints. Similarly, to maximize sum rate, \cite{Chen2017_LowComplexityBeamformingUserSelection} studies the channel state information based singular value decomposition precoding scheme. Total power minimization with QoS requirements and total power minimization under target interference level constraints are respectively investigated in 
\cite{Alavi2018_BeamformingTechniquesforNOMA} and \cite{Chen2016_OptimalPrecodingForQOSOpt}. In addition, a max-min fair (MMF) precoder design problem for a multiple antenna base station is also studied in \cite{Alavi2018_BeamformingTechniquesforNOMA}. Power allocation (PA) problems for achieving MMF in NOMA systems with single antenna transmitters are studied in \cite{Timotheou2015} and \cite{choi2016}.

As mentioned above, in NOMA, a single precoder vector can be shared by a cluster of users. For this case, weighted sum rate optimization under a total power constraint when two users exist in each cluster is studied in \cite{XSun2015_NOMAWithWSR}. For clustered downlink NOMA systems, a sub-optimal user clustering algorithm is proposed and the optimal power allocation policy that maximizes the weighted sum rate is derived in \cite{Ali2016_DynamicUser}, \cite{Ali2017_NOMAForDownlinkMUMIMOSystems}. Joint power allocation and precoder design to maximize the strong users' sum rate subject to QoS constraints on weak users' rates is solved via successive convex approximation (SCA) and semi-definite relaxation (SDR) in \cite{XSun2016_JointBeamformingandPA}. The same problem is generalized to the multi-cell networks in \cite{XSun2018_JointBeamformingandPA}. Finally, minimizing total transmission power for downlink clustered NOMA is studied in \cite{Choi2015_MinimumPowerMulticastAndNOMA} and \cite{Liu2017_JointBeamformingPowerOpt}. 



\vspace{-0.4cm}
\subsection{Motivation and Contributions}

In this work, we study downlink MIMO clustered NOMA system from a fairness standpoint and we investigate joint precoder design and PA problem that provide MMF among the strongest users in each cluster and ensure the minimum rate requirements for all the other users. To the best of our knowledge, there is no joint MMF precoder design and PA optimization for a clustered downlink NOMA system. Our contributions are listed below:
\begin{enumerate}
\item Firstly, we define a joint precoder design and PA problem to attain max-min fairness among the best users in each cluster, while guaranteeing target data rates for the rest of the users. Due to the non-convexity of the defined problem, we apply Taylor series expansion, SDR, to simplify the original problem. Next, we propose a suboptimal iterative SCA based algorithm. 
\item Secondly, we use the equivalency between weighted mean square error (WMMSE) and achievable rate expressions, and restate the original problem as an equivalent MMF WMMSE problem. To do that, we apply the achievable rate-WMMSE relationship. Due to the non-convexity of the main problem, we split it into two different problems: i) to design optimal precoders for given power allocation coefficients (PAC), and ii) to obtain optimal PAC for given precoders. We derive a sub-optimal PA scheme while designing the optimal precoders. Employing the CVX toolbox to obtain the precoders, we then propose a suboptimal iterative WMMSE based algorithm, which updates transmit precoders, receivers, weights and PAC sequentially.
\item Thirdly, employing the Karush-Kuhn-Tucker (KKT) optimality conditions, we find the expressions the optimal receivers, MMSE weights and the optimal precoders have to satisfy. Utilizing these expressions, we propose a low-complexity iterative algorithm to evaluate precoders and receivers. We use the exponential penalty method to evaluate the Lagrange multipliers. We find that this approach significantly decreases complexity, while ensuring a similar MMF rate performance as the CVX solution in the second item.
\item To the best of our knowledge, there is no work in the literature, which studies both SDR/SCA and WMMSE based approaches for the same optimization problem. We discover that SDR/SCA performs better than the latter as it solves a tighter approximation.
\item We compare the proposed schemes with and without power allocation to observe that power optimization does not significantly increase complexity and its advantages in terms of MMF rates are justified.
\item We also compare our results with rate-splitting (RS) \cite{Mao2018, Mao2019}, OMA and multi-user linear precoding (MULP) schemes in terms of MMF rates, complexity and energy efficiency. Our results reveal that the SDR/SCA based scheme offers an excellent tradeoff in all three aspects.
\end{enumerate}

Next, we explain the system model and define the optimization problem in Section \ref{sec:systemmodel}. We propose the SDR/SCA and WMMSE based precoder designs respectively in Sections \ref{sec:SCA} and \ref{sec:WMMSE}. We present the numerical results in Section \ref{sec:sim}. Finally we provide conclusions and future work in Section \ref{sec:conc}.

\section{System Model and Problem Definitions}\label{sec:systemmodel}

In this paper, we investigate a downlink multiuser MIMO system. The base station has $M$ transmit antennas and communicates with $K$ clusters. There are $L$ single antenna users in each cluster\footnotemark\footnotetext{In fact, the results can easily be extended to cover for unequal number of users in each group. However, to keep the notation simple we adhere to a fixed number of users in each cluster.} and each user belongs to only one cluster. 

The base station aims to send the data $s_{k,l}$ to the $l$-th user in the $k$-th cluster, for all $l \in \{1,\ldots,L\}$, and $k \in \{1,\ldots,K\}$. All $s_{k,l}$ are independent and $\mathbb{E}\{{s}_{k,l}{s}_{k,l}^{\ast}\} = \alpha_{k,l}$. Here $\alpha_{k,l}$ is the ratio of power allocated to the data stream $s_{k,l}$. The PAC vector is defined as $\mathbf{A} = [\alpha_{1,1},\ldots,\alpha_{1,L}, \ldots,\alpha_{K,1},\ldots, \alpha_{K,L}]$. Moreover, $\sum_{l=1}^{L} \alpha_{k,l} = 1$. To send all the messages, the base station superposes all the messages in a cluster as $s_k = \sum_{l=1}^{L} s_{k,l}$ and forms $\mathbf{s} = {[{s}_1,\ldots,{s}_{K}]}^T$ $\in \mathbb{C}^{K \times 1}$. When $\mathbf{p}_k \in \mathbb{C}^{M\times 1}$ indicates the precoder vector for the $k$-th cluster, the base station transforms $\mathbf{s}$ with the precoder matrix $\mathbf{P} = [\mathbf{p}_1,\ldots,\mathbf{p}_{K}]$  $\in \mathbb{C}^{M \times K}$. Then, the base station transmits $\mathbf{x}$ $\in \mathbb{C}^{M\times 1}$, which is equal to
\begin{align}
\mathbf{x} = \mathbf{P}\mathbf{s} &= \sum_{k=1}^{K} \mathbf{p}_k {s}_{k}=\sum_{k=1}^{K} \sum_{l=1}^{L} \mathbf{p}_k {s}_{k,l}.\label{precodedSignal1}
\end{align}
The base station has an average total power constraint $E_{tx}$, which is written as
\begin{align}
\mathbb{E}\{{\mathbf{x}}^{H}{\mathbf{x}}\}&= \Tr(\mathbf{P}{\mathbf{P}}^H )\leq E_{tx}. \label{pow_const}
\end{align} Then, the received signal at the $l$-th user in the $k$-th cluster becomes
\begin{align}
 y_{k,l} &= \mathbf{h}_{k,l}^{H} \mathbf{p}_k  \sum_{l=1}^{L} s_{k,l} + \mathbf{h}_{k,l}^{H} \sum_{i=1,i\neq k}^{K} \mathbf{p}_i  s_{i} + n_{k,l}.\label{rec_signal}
\end{align}Here, ${\mathbf{h}_{k,l}}$ $\in \mathbb{C}^{M \times 1}$ is the effective channel gain vector of the $l$-th user in the $k$-th cluster. The effective channel gain is defined as ${\mathbf{h}_{k,l}} = {\mathbf{\tilde{h}}}_{k,l} /\sqrt{d_{k,l}^{\rho}}$, where $d_{k,l}$ is the distance between the $l$-th user in the $k$-th cluster and the base station, and $\rho$ is the path loss exponent. The entries in $\mathbf{\tilde{h}}_{k,l}$ are independent and identically distributed (i.i.d.) and complex valued random variables. 
Moreover, the effective channel gain magnitudes are ordered as $|h_{k,L}| > |h_{k,L-1}| > \ldots > |h_{k,1}|$. It means that the user with the smallest effective channel gain magnitude is the first user in a cluster and the $L$-th user has the largest channel gain magnitude. The noise component $n_{k,l}$ is a circularly symmetric complex Gaussian random variable with zero mean and unit variance, and $n_{k,l}$ are i.i.d. for all $k$ and $l$. The base station is informed about all effective channel gains $\mathbf{h}_{k,l}$, while the receivers know only their own $\mathbf{h}_{k,l}$.


\subsection{Achievable Data Rates}
For this NOMA system we investigate,  the messages for different clusters will be treated as noise, while SIC will be carried out within a cluster to limit intra-cluster interference. Due to SIC, in the $k$-th cluster, the $l$-th user's message is decoded at the $i$-th user, for which $l \leq i$. In other words, decoding is ordered and starts from the first user's message. The first user in the cluster decodes its own message only, and the $L$-th user decodes all users' messages within the cluster. To simplify the notation, we define the sets $\mathcal{K} \triangleq  \{1, . . . , K\}$, $\mathcal{L} \triangleq  \{1, . . . , L\}$, $\bar{\mathcal{L}} \triangleq  \{1, . . . , L-1\}$ and $\mathcal{I} \triangleq  \{l, . . . , L\}$. Then, the signal to interference ratio (SINR) for decoding the $l$-th user's message at the $i$-th user in the $k$-th cluster, $i \in \mathcal{I}, l \in \mathcal{L}, k\in\mathcal{K}$ can be written as
\begin{align}
\gamma_{k, i \rightarrow l} &= \alpha_{k,l} |\mathbf{h}_{k,i}^{H} \mathbf{p}_k|^2{ r_{k, i \rightarrow l}^{-1}} \label{SINR_1}.
\end{align}
In the above equation, $r_{k, i \rightarrow l}$ is the effective noise variance and is defined as
\begin{align}
r_{k, i \rightarrow l} &= \sum_{j = l+1 }^{L}\alpha_{k,j} |\mathbf{h}_{k,i}^{H} \mathbf{p}_k|^2  + \sum_{t =1 ,t\neq k}^{K} |\mathbf{h}_{k,i}^{H} \mathbf{p}_{t}|^2  + 1.
\end{align} Then, in the $k$-th cluster, the $i$-th user's achievable rate\footnotemark\footnotetext{In all the derivations, all rate expressions are expressed in nats/channel use. In Section 
\ref{sec:sim}, without loss of generality, simulation results are presented in bits/channel use.} for decoding the $l$-th user's message is 
\begin{align}
R_{k, i \rightarrow l} &= 
 \log \left( {1} + \gamma_{k, i \rightarrow l} \right). \label{rate_u_m}
\end{align} Overall, the achievable rate for the $l$-th user's message in the $k$-th cluster is defined as the minimum of all $R_{k, i \rightarrow l}$, and is denoted as
\begin{align}
R_{k,l} &= \min_{i, i \in \mathcal{I}}R_{k, i \rightarrow l}, \forall l \in \bar{\mathcal{L}}
\label{rate_u_min}.
\end{align} Note that, due to this definition, $R_{k,L} = R_{k, L \rightarrow L}$.

\subsection{Max-Min Fair Problem Definition}\label{subsec:probdef}

In this subsection, we define the MMF rate optimization problem, which aims to find the optimal precoder matrix $\mathbf{P}$ and optimal PAC vector $\mathbf{A}$, such that the minimum of the strongest users' rates 
is maximized subject to a total power constraint and a minimum rate constraint for the rest of the users. Then, the optimization problem is stated as
\begin{subequations}\label{MMF_WSR1_all}
\begin{align}
\max_{\mathbf{P},\mathbf{A}} &\qquad \min_{k\in \mathcal{K}} \:    R_{k, L}  \label{MMF_WSR1} \\
 \text{s.t. }
 &\qquad  R_{k,l}^{th} \leq  R_{k, i \rightarrow l}, \forall k \in \mathcal{K}, \forall l \in \bar{\mathcal{L}}, \forall i \in \mathcal{I}, \label{MMF_WSR_cnst1} \\
&\qquad  \sum_{l=1}^{L} \alpha_{k,l}= 1,  \alpha_{k,l} \geq 0,  \;\forall k \in \mathcal{K}, \forall l \in \mathcal{L}, \label{MMF_WSR_cnst2}\\
 &\qquad  \Tr (\mathbf{P}\mathbf{P}^H) \leq E_{tx}, \label{MMF_WSR_powCnst}
\end{align}
\end{subequations}where $R_{k,l}^{th} \geq 0$ is the threshold data rate that has to be provided to the $l$-th user in the $k$-th cluster $\forall k \in \mathcal{K}, \forall l \in \bar{\mathcal{L}}$. 
Note that, due to SIC, the $l$-th user's message in the $k$-th cluster has to be decoded by all $i$, $\forall i \in \mathcal{I}$, resulting in the inequality in \eqref{MMF_WSR_cnst1}. 
The equality in \eqref{MMF_WSR_cnst2} indicates that the superposed data $s_k$ for the $k$-th cluster has normalized power. In addition, \eqref{MMF_WSR_powCnst} is the total power constraint at the base station.

To solve this problem, we need to restate $\eqref{MMF_WSR1_all}$, as the minimum operation in the objective function is not a convex function. Thus, we add an auxiliary variable $R_g$ and convert $\eqref{MMF_WSR1_all}$ to a new constrained optimization problem as
\begin{subequations}\label{MMF_WSR2_all}
\begin{align}
\max_{\mathbf{P},\mathbf{A}, R_g} &\qquad R_g \label{MMF_WSR2} \\
 \text{s.t. }&\qquad  R_g \leq  R_{k,L}, \forall k \in \mathcal{K}, \label{MMF_WSR2_cnst1} \\
 &\qquad  \eqref{MMF_WSR_cnst1}, \eqref{MMF_WSR_cnst2}, \eqref{MMF_WSR_powCnst}.
\end{align}
\end{subequations}

\section{Successive Convex Approximation Solution} \label{sec:SCA}

The problem defined in \eqref{MMF_WSR2_all} is still a non-convex optimization problem. In this section, we further modify the optimization problem in \eqref{MMF_WSR2_all} to obtain an equivalent semi-definite programming problem. 

To achieve this objective, we introduce and optimize the auxiliary optimization matrix $\mathbf{Q}_k = \mathbf{p}_k \mathbf{p}_k^{H}$. Note that, $\mathbf{Q}_k \in \mathbb{C}^{M \times M}$ is a rank-one positive semi-definite matrix. Then, we can rewrite our optimization problem as
\begin{subequations}\label{MMF_WSR3_all}
\begin{align}
\max_{\substack{\mathbf{Q}, \mathbf{A},\\ R_g}} &\quad R_g \label{MMF_WSR3} \\
 \text{s.t. }
 &\quad  R_g \leq  \log \left( 1 +  \tilde{\gamma}_{k,L}  \right), \forall k \in \mathcal{K}, \label{MMF_WSR3_cnst1} \\
 &\quad R_{k,l}^{th} \leq \log \left(1 + \tilde{\gamma}_{k, i \rightarrow l}\right), \forall k \in \mathcal{K}, \forall l \in \bar{\mathcal{L}}, \forall i \in \mathcal{I}, \label{MMF_WSR3_cnst2}\\
&\quad  \sum_{l=1}^{L} \alpha_{k,l}= 1,  \alpha_{k,l} \geq 0,  \;\forall k \in \mathcal{K}, \forall l \in \mathcal{L}, \label{MMF_WSR3_cnst3}\\
&\quad  \mathbf{Q}_k \succeq 0, \forall k \in \mathcal{K} \label{MMF_WSR3_cnst4},\\
&\quad \mathrm{rank}\left(\mathbf{Q}_k\right) \leq 1, \forall k \in \mathcal{K} \label{MMF_WSR3_cnst5},\\
&\quad \sum_{k=1}^{K} \Tr \left(\mathbf{Q}_k\right) \leq E_{tx},\label{MMF_WSR3_cnst6}
\end{align}
\end{subequations}
where
\begin{align}
\tilde{\gamma}_{k,L} &= \frac{\alpha_{k,L} \mathbf{h}_{k,L}^{H} \mathbf{Q}_k \mathbf{h}_{k,L}} {\sum_{t=1,t \neq k}^{K} \mathbf{h}_{k,L}^{H} \mathbf{Q}_t \mathbf{h}_{k,L} + 1},\\
\tilde{\gamma}_{k, i \rightarrow l} &= \nonumber \\
&\frac{\alpha_{k,l} \mathbf{h}_{k,i}^{H} \mathbf{Q}_k \mathbf{h}_{k,i}} {\sum_{j=l+1}^{L} \alpha_{k,j} \mathbf{h}_{k,i}^{H} \mathbf{Q}_k \mathbf{h}_{k,i} + \sum_{t=1,t \neq k}^{K} \mathbf{h}_{k,i}^{H} \mathbf{Q}_t \mathbf{h}_{k,i} + 1 }.
\end{align}

Convex optimization solvers are not efficient when operating with logarithmic functions. To eliminate the logarithms in \eqref{MMF_WSR3_cnst1} and \eqref{MMF_WSR3_cnst2}, we define a new auxiliary variable $\delta$ and new constants $\zeta_{k,l}$, $\forall k \in \mathcal{K}, \forall l \in \bar{\mathcal{L}}$ as 
\begin{eqnarray*}
\delta &=& e^{R_g} - 1 \\
\zeta_{k,l}& =& e^{R_{k,l}^{th}} - 1.
\end{eqnarray*}
Then, we can reformulate \eqref{MMF_WSR3_all} as
\begin{subequations}\label{MMF_WSR4_all}
\begin{align}
\max_{\substack{\mathbf{Q}, \mathbf{A}, \delta}} &\quad \delta \label{MMF_WSR4} \\
 \text{s.t. }
 &\quad  \delta \leq  \frac{\alpha_{k,L} \phi_{k,L}}{\omega_{k,L}}, ~\forall k \in \mathcal{K}, \label{MMF_WSR4_cnst1} \\
 &\quad \zeta_{k,l}  \leq \frac{\mu_{k,l}\phi_{k,i}}{\omega_{k,i}} ~~\forall k \in \mathcal{K}, \forall l \in \bar{\mathcal{L}}, \forall i \in \mathcal{I}, \label{MMF_WSR4_cnst2}\\
&\quad \eqref{MMF_WSR3_cnst3}, \eqref{MMF_WSR3_cnst4}, \eqref{MMF_WSR3_cnst5}, \eqref{MMF_WSR3_cnst6},
\end{align}
\end{subequations}
where
\begin{align}
    \phi_{k,l} &= \mathbf{h}_{k,l}^{H}\mathbf{Q}_k \mathbf{h}_{k,l} \label{eq:phi_kl}\\
    \omega_{k,i} &= 1+ \sum_{t=1,t \neq k}^{K} \mathbf{h}_{k,i}^{H}\mathbf{Q}_t \mathbf{h}_{k,i}\label{eq:omega_ki}\\
    \mu_{k,l} &= \alpha_{k,l} - \zeta_{k,l} \sum_{j=l+1}^{L} \alpha_{k,j} .\label{eq:mu_kl}
\end{align} The constraints \eqref{MMF_WSR4_cnst1} and \eqref{MMF_WSR4_cnst2} are not convex, since $\alpha_{k,L}\phi_{k,L}$ and $\mu_{k,l}\phi_{k,i}$ are both bilinear functions. To change \eqref{MMF_WSR4_cnst1} and \eqref{MMF_WSR4_cnst2} into convex constraints, we need to apply the Schur complement \cite{ShurComplement}. Introducing new auxiliary variables $\tau_{k,i,l}$, we can replace \eqref{MMF_WSR4_cnst1} and \eqref{MMF_WSR4_cnst2} with the following 4 new constraints
\begin{align}
    \begin{bmatrix}
        \alpha_{k,L} & \tau_{k,L,L}\\
        \tau_{k,L,L} & \phi_{k,L}
    \end{bmatrix} &\succeq \mathbf{0} , ~\forall k \in \mathcal{K} \\
    \begin{bmatrix}
        \mu_{k,l} & \tau_{k,i,l}\\
        \tau_{k,i,l} & \phi_{k,i}
    \end{bmatrix} &\succeq \mathbf{0} ,~\forall k \in \mathcal{K}, \forall l \in \bar{\mathcal{L}}, \forall i \in \mathcal{I} 
    \end{align}
and
   \begin{align}
    \delta &\leq \frac{\tau_{k,L,L}^2}{\omega_{k,L}} ,~ \forall k \in \mathcal{K} \label{eq:dcoptfun}\\
    \zeta_{k,l}  &\leq \frac{\tau_{k,i,l}^2}{\omega_{k,i}},~\forall k \in \mathcal{K},~ \forall l \in \bar{\mathcal{L}},~ \forall i \in \mathcal{I}. \label{eq:cthrfun}
\end{align}
The right-hand side of \eqref{eq:dcoptfun} is convex in both $\tau_{k,L,L}$ and $\omega_{k,L}$, and the right-hand side of \eqref{eq:cthrfun} is convex in both $\tau_{k,i,l}$ and $\omega_{k,i}$. In other words, right hand sides of both \eqref{eq:dcoptfun} and \eqref{eq:cthrfun} are difference-of-convex functions \cite{Khabbazibasmenj2012}. Therefore, we can apply the first-order Taylor expansions \cite{TaylorSeries} to obtain a tight lower bound on these two functions. 
For given fixed points $(\tilde{\tau}_{k,L,L}$, $\tilde{\omega}_{k,L})$ $\forall k \in \mathcal{K}$ and $(\tilde{\tau}_{k,i,l}, \tilde{\omega}_{k,i}), \forall k \in \mathcal{K}, \forall l \in \bar{\mathcal{L}}, \forall i \in \mathcal{I}$, we write
\begin{align}
     \delta &\leq \frac{2\tilde{\tau}_{k,L,L}}{\tilde{\omega}_{k,L}} \tau_{k,L,L} - \frac{\tilde{\tau}_{k,L,L}^2}{\tilde{\omega}_{k,L}^{2}}  \omega_{k,L}
     \leq \frac{\tau_{k,L,L}^2}{\omega_{k,L}}\\
 \zeta_{k,l}  & \leq \frac{2 \tilde{\tau}_{k,i,l}}{\tilde{\omega}_{k,i}} \tau_{k,i,l} - \frac{\tilde{\tau}_{k,i,l}^2}{\tilde{\omega}_{k,i}^2} \omega_{k,i}\leq \frac{\tau_{k,i,l}^2}{\omega_{k,i}}
\end{align}where $\tilde{\tau}_{k,i,l} \geq 0$ and $\tilde{\omega}_{k,i} \geq 1$. 

Finally, we relax the equality in \eqref{MMF_WSR3_cnst3} as an inequality, omit the constraint in \eqref{MMF_WSR3_cnst5} and transform the optimization defined in \eqref{MMF_WSR3_all} as
\vspace{-0.7cm}
\begin{subequations}\label{MMF_WSR5_all}
\begin{align}
\max_{\substack{\mathbf{Q}, \mathbf{A}, \delta, \tau}} &\quad \delta \label{MMF_WSR5} \\
 \text{s.t. }
 &\quad \delta \leq \frac{2\tilde{\tau}_{k,L,L}}{\tilde{\omega}_{k,L}} \tau_{k,L,L} - \frac{\tilde{\tau}_{k,L,L}^2}{\tilde{\omega}_{k,L}^{2}}  \omega_{k,L}, \forall k \in \mathcal{K}, \label{MMF_WSR5_cnst1}\\
 & \quad \zeta_{k,l}  \leq \frac{2 \tilde{\tau}_{k,i,l}}{\tilde{\omega}_{k,i}} \tau_{k,i,l} - \frac{\tilde{\tau}_{k,i,l}^2}{\tilde{\omega}_{k,i}^2} \omega_{k,i}, \nonumber \\
 &~~~~~~~~~~~~~~~~~~~~~~~~~~~\forall k \in \mathcal{K}, \forall l \in \bar{\mathcal{L}}, \forall i \in \mathcal{I},\label{MMF_WSR5_cnst2} \\
      & \quad \begin{bmatrix}
        \alpha_{k,L} \quad \tau_{k,L,L}\\
        \tau_{k,L,L} \quad \phi_{k,L}
    \end{bmatrix} \: \succeq \mathbf{0} , \forall k \in \mathcal{K}, \label{MMF_WSR5_const3}\\
    & \quad \begin{bmatrix}
        \mu_{k,l} \quad  \tau_{k,i,l}\\
        \tau_{k,i,l} \quad  \phi_{k,i}
    \end{bmatrix} \: \succeq \mathbf{0} ,\forall k \in \mathcal{K}, \forall l \in \bar{\mathcal{L}}, \forall i \in \mathcal{I}, \label{MMF_WSR5_const4} \\
&\quad  \sum_{l=1}^{L} \alpha_{k,l}= 1,  \alpha_{k,l} \geq 0,  \;\forall k \in \mathcal{K}, \forall l \in \mathcal{L}, \label{MMF_WSR5_cnst5}\\
&\quad  \mathbf{Q}_k \succeq 0, \forall k \in \mathcal{K}, \label{MMF_WSR5_cnst6}\\
&\quad \sum_{k=1}^{K} \Tr \left(\mathbf{Q}_k\right) \leq E_{tx}.\label{MMF_WSR5_cnst7}
\end{align}
\end{subequations}

The problem in \eqref{MMF_WSR5_all} is a constrained convex optimization problem when $\tilde{\tau}_{k,i,l}$ and $\tilde{\omega}_{k,i}$ are given. In \cite{XSun2018_JointBeamformingandPA}, the authors prove that omitting the rank constraint \eqref{MMF_WSR3_cnst5} in \eqref{MMF_WSR5_all} does not alter the problem. They discuss that the solution is always rank one. However, their proof assumes that the principle eigenvalue of the positive semi-definite matrix in \cite[eqn. (33)]{XSun2018_JointBeamformingandPA} is always unique. This may not be the case and there can be more than one principle eigenvalue. However, adding independent rank one matrices in \cite[eqn. (33)]{XSun2018_JointBeamformingandPA} results in full rank matrices with very high probability and this does not pose a significant issue.

The optimization problem \eqref{MMF_WSR5_all} is an approximation to the original problem in  \eqref{MMF_WSR1_all}. To solve \eqref{MMF_WSR1_all}, we use Algorithm \ref{algorithm:algo_SCA}. The algorithm solves \eqref{MMF_WSR5_all} when $\tilde{\tau}_{k,i,l}$ and $\tilde{\omega}_{k,i}$ are given, and updates these values in each iteration. While solving \eqref{MMF_WSR5_all}, we employ the CVX optimization toolbox \cite{CVX}.
\begin{algorithm}[!t]
    \caption{SDR/SCA Based MMF Algorithm with PA}
    \label{algorithm:algo_SCA}
    \begin{algorithmic}[1]
    \State \textbf{Input:} $\mathbf{A}^{(0)}$, $E_{tx}$, $\Upsilon$, $R_{k,l}^{th}$, $n_{max}$ \\
    \textbf{Initialize:} $\delta^{(0)}=0$, $\tilde{\omega}_{k,i}^{(0)}, \tilde{\tau}_{k,i,l}^{(0)}$, and $n=0$;\\
    \textbf{iterate} $\forall j, l, k$;
      \State \quad $n = n +1$ 
      \State \quad \multiline{Update $\left\{\mathbf{Q}_k^{(n)}, \mathbf{A}^{(n)}, \delta^{(n)}, \tau_{k,i,l}^{(n)}  
      \right\}$ by solving \eqref{MMF_WSR5_all} \\for given $\tilde{\tau}_{k,i,l}^{(n-1)}$ and $\tilde{\omega}_{k,i}^{(n-1)}$}
      \State \quad Update $\tilde{\tau}_{k,i,l}^{(n)}$ and $\tilde{\omega}_{k,i}^{(n)}$ using \eqref{eq:iterVar}
      \State \quad $\mathbf{If}$ {$ (\delta^{(n)} - \delta^{(n-1)})/\delta^{(n-1)}  < \Upsilon $} $\textbf{or}$ $n=n_{max}$ $\mathbf{then}$
      \State \quad \quad Terminate 
      \State \quad $\mathbf{else}$ $\mathbf{then}$ 
      \State \quad \quad Go to Step 3
    \end{algorithmic}
\end{algorithm} 

In Algorithm \ref{algorithm:algo_SCA}, we can initialize $\tilde{\tau}_{k,i,l}^{(0)}$ and $\tilde{\omega}_{k,i}^{(0)}$ arbitrarily, as long as $\tilde{\tau}_{k,i,l}^{(0)} \geq 0$ and $\tilde{\omega}_{k,i}^{(0)} \geq 1$. However, one can initialize $\tilde{\tau}_{k,i,l}^{(0)}$ and $\tilde{\omega}_{k,i}^{(0)}$ more efficiently. To do so, we create a random rank one positive semi-definite matrix for each $\mathbf{Q}_k$ and a uniform vector $\mathbf{A}$, calculate $\phi_{k,l}^{(0)}$ and $\omega_{k,i}^{(0)}$ and compute $\tilde{\tau}_{k,i,l}^{(0)}$ and $\tilde{\omega}_{k,i}^{(0)}$ as
\begin{align}\label{eq:tauAndomegaZero}
       \tilde{\tau}_{k,i,l}^{(0)} =
      \begin{cases} 
      \sqrt{\alpha_{k,L} \phi_{k,L}^{(0)}}, & i=l=L \\
      \sqrt{\mu_{k,l} \phi_{k,i}^{(0)}}, & l \leq i, l<L 
   \end{cases}\;\;\;\text{and}\;\;\; \tilde{\omega}_{k,i}^{(0)} = \omega_{k,i}^{(0)} 
\end{align}
In \eqref{eq:tauAndomegaZero}, the $\tilde{\tau}_{k,i,l}^{(0)}$ values satisfy \eqref{MMF_WSR5_const3} and \eqref{MMF_WSR5_const4} with equality. Using the randomly generated matrices for $\mathbf{Q}_k$, we calculate $\tilde{\omega}_{k,i}^{(0)}$ using \eqref{eq:omega_ki}. After initialization, in Algorithm \ref{algorithm:algo_SCA}, we update $\tilde{\tau}_{k,i,l}$ and $\tilde{\omega}_{k,i}$ in each iteration as
\begin{align}
    \tilde{\tau}_{k,i,l}^{(n)} = \tau_{k,i,l}^{(n-1)} \;\;\; \text{and}\;\;\;   \tilde{\omega}_{k,i}^{(n)} = \omega_{k,i}^{(n-1)}. \label{eq:iterVar}
\end{align} This way, the bounds in \eqref{MMF_WSR5_cnst1} and \eqref{MMF_WSR5_cnst2} become tighter in each iteration. The algorithm convergence can be proved in a similar manner as in \cite{XSun2018_JointBeamformingandPA}.

\section{WMMSE Based Solutions}\label{sec:WMMSE}

In this section, we provide an alternative solution to the MMF problem defined in \eqref{MMF_WSR1_all} using the MMSE approach. In this approach, we utilize the relation between mutual information and MMSE \cite{mutual_mmse,Christensen2008_WeightedSumRate}. We can state $R_{k, i \rightarrow l}$ in terms of error variances, assuming MMSE receivers are
employed at the receivers. 

We remind that the effective channel gain magnitudes are ordered as $|h_{k,L}|>|h_{k,L-1}|> \ldots |h_k,1|$ as described in Section \ref{sec:systemmodel}. Therefore, the $l$-th user in the $k$-th cluster decodes messages in order starting from the first user's message, and decodes its own message in the last step. SIC is employed in each step. In other words, to estimate the $l$-th user's message, the $i$-th user ($i \in \mathcal{I}$) in the $k$-th cluster employs the SIC receiver ${V}_{k, i \rightarrow l}$ on its equivalent received signal $\hat{y}_{k,i}$ where 
\begin{eqnarray} \hat{y}_{k,i} = {y}_{k,i} - \mathbf{h}_{k,i}^{H}\mathbf{p}_{k} \sum_{
j = 1}^{l-1} s_{k,j}. \end{eqnarray}
The $i$-th user's estimate $\hat{{s}}_{k, i \rightarrow l}$ about $s_{k,l}$ becomes
\begin{align}
\hat{{s}}_{k, i \rightarrow l} = {V}_{k, i \rightarrow l}\hat{y}_{k,i}.
\end{align}
Then, the MSE of the $i$-th user's estimate of the $l$-th user's message in the $k$-th cluster can be written as 
\begin{align}
\varepsilon_{k, i \rightarrow l} &= \mathbb{E}\big\{||\hat{{s}}_{k, i \rightarrow l} - s_{k,l}||^2\big\} \nonumber \\
&=|{V}_{k, i \rightarrow l}|^2 T_{k, i \rightarrow l} + \alpha_{k,l} - 2 \mathfrak{R}\left\{\alpha_{k,l} {V}_{k, i \rightarrow l} \mathbf{h}_{k,i}^{H} \mathbf{p}_{k} \right\}, \label{UMSE_k1}
\end{align}where $T_{k, i \rightarrow l} = |_{k,i}\mathbf{p}_{k}|^2 \alpha_{k,l} + r_{k, i \rightarrow l}$.
Given above, the optimal MMSE receiver is
\begin{align}
V_{k, i \rightarrow l}^{mmse} &= \arg\min_{{V}_{k, i \rightarrow l}} \varepsilon_{k, i \rightarrow l} 
 = \alpha_{k,l} \mathbf{p}_k^{H} \mathbf{h}_{k,i} T_{k, i \rightarrow l}^{-1}\label{V_rec1}.
\end{align}When this MMSE receiver in (\ref{V_rec1}) is employed, the resulting error variance expression in (\ref{UMSE_k1}) becomes
\begin{align}
\varepsilon_{k, i \rightarrow l}^{mmse}  &=\left( \frac{1}{\alpha_{k,l}} +  |\mathbf{h}_{k,i}^{H} \mathbf{p}_k|^2 r_{k, i \rightarrow l}^{-1}  \right)^{-1}. \label{error_cov1}
\end{align} 
As the message for the $l$-th user has to be decoded by all users $i$ for which $i \geq l$ in the $k$-th cluster, we define $\varepsilon_{k,l}^{mmse}$ as
\begin{align}
\varepsilon_{k,l}^{mmse} &= \max_{i, i \in \{l,...,L\}}{\varepsilon}_{k, i \rightarrow l}^{mmse}. \label{mmse_u_min}
\end{align} Note that, by simply comparing the rate and MMSE expressions in (\ref{rate_u_m}) and (\ref{error_cov1}) we observe that
\begin{align}
R_{k, i \rightarrow l} &= \log \left[\alpha_{k,l}{\varepsilon}_{k, i \rightarrow l}^{mmse^{-1}}\right].\label{rate_errrorCov1}
\end{align}

\subsection{Equivalent MMF WMMSE Problem}

To convert \eqref{MMF_WSR2_all} into an equivalent WMMSE problem, we use the above relation between rate and MMSE. We define the augmented weighted MSE \cite{Christensen2008_WeightedSumRate} as
\begin{align}
\xi_{k, i \rightarrow l} &= b_{k, i \rightarrow l} \varepsilon_{k, i \rightarrow l} - \log (\alpha_{k,l} b_{k, i \rightarrow l} ) ,\label{ksi_u}
\end{align}where $b_{k, i \rightarrow l}> 0$ is the weight for MSE. We also define the minimum of the augmented WMSEs as
\begin{align}
\xi_{k, i \rightarrow l}^{mmse} & \triangleq  \arg \min_{\{{b}_{k, i \rightarrow l}, {V}_{k, i \rightarrow l}\}} \xi_{k, i \rightarrow l},\\
& = b_{k, i \rightarrow l}^{mmse} \varepsilon_{k, i \rightarrow l}^{mmse} - \log (\alpha_{k,l} b_{k, i \rightarrow l}^{mmse}). \label{eqn:2}
\end{align}
It is seen that the augmented WMSE $\xi_{k, i \rightarrow l}$ is convex in the receiver $V_{k, i \rightarrow l}$. Solving for the first order optimality conditions in \eqref{ksi_u}, we find the optimum receiver in \eqref{eqn:2} as ${V_{k, i \rightarrow l}^\star} = V_{k, i \rightarrow l}^{mmse}$ and the optimum weights as
\begin{align}
b_{k, i \rightarrow l}^\star &= b_{k, i \rightarrow l}^{mmse} =\frac{1}{\varepsilon_{k, i \rightarrow l}^{mmse}}, \label{b_weight}
\end{align}where the MMSE receiver $V_{k, i \rightarrow l}^{mmse}$ is given in \eqref{V_rec1} and the MMSE error variance $\varepsilon_{k, i \rightarrow l}^{mmse}$ is given in \eqref{error_cov1}.

One can obtain the relation between rate expressions and augmented WMSEs by checking the first order optimality conditions \cite{Christensen2008_WeightedSumRate} to find that
\begin{align}
\xi_{k, i \rightarrow l}^{mmse} &= 1 - R_{k, i \rightarrow l}. \label{ksi_kil_MMSE}
\end{align}

Utilizing the equality in (\ref{ksi_kil_MMSE}), the optimization problem in (\ref{MMF_WSR2_all}) can be written as:
\begin{subequations}\label{MMF_MSE1_all}
\begin{align}
\max_{\substack{\mathbf{P}, \mathbf{A},\\ \mathbf{R}, R_g}} &\quad R_g \label{MMF_MSE1} \\
 \text{s.t. }
&\quad  R_g \leq R_{k,L}, \forall k \in \mathcal{K},\label{MMF_MSE1_cnst1}\\
&\quad  R_{k,l}^{th} \leq R_{k,l}, \forall k \in \mathcal{K}, \forall l \in \bar{\mathcal{L}}, \label{MMF_MSE1_cnst2}\\
&\quad  R_{k,l} \leq  1 - \xi_{k, i \rightarrow l}^{mmse}, \forall k \in \mathcal{K}, \forall l \in \mathcal{L}, \forall i \in \mathcal{I}, \label{MMF_MSE1_cnst3} \\
&\quad  \eqref{MMF_WSR_cnst2}, \eqref{MMF_WSR_powCnst} \label{MMF_MSE1_cnst4},
\end{align}
\end{subequations} where $\mathbf{R} = [R_{1,1},\ldots,R_{1,L}, \ldots,R_{K,1},\ldots, R_{K,L}]$ is a new auxiliary variable vector.

The optimization problem in \eqref{MMF_MSE1_all} assumes that the optimal MMSE receiver defined in (\ref{V_rec1}) is employed at all users, and finds the optimal precoders at the transmitter. Below, we first define a generalized problem which allows for arbitrary receivers ${V}_{k, i \rightarrow l}$ that attain $\varepsilon_{k, i \rightarrow l}$ in (\ref{UMSE_k1}).
\begin{subequations}\label{MMF_MSE11_all}
\begin{align}
\max_{\substack{\mathbf{P}, \mathbf{A}, \mathbf{R},\\ R_g, \mathbf{V}}} &\quad R_g \label{MMF_MSE11} \\
 \text{s.t. }
&\quad  R_g \leq R_{k,L}, \forall k \in \mathcal{K}, \label{MMF_MSE11_cnst1}\\
&\quad  R_{k,l}^{th} \leq R_{k,l}, \forall k \in \mathcal{K}, \forall l \in \bar{\mathcal{L}}, \label{MMF_MSE11_cnst2}\\
&\quad  R_{k,l} \leq  1 - \xi_{k, i \rightarrow l}, \forall k \in \mathcal{K}, \forall l \in \mathcal{L}, \forall i \in \mathcal{I}, \label{MMF_MSE11_cnst3} \\
&\quad  \eqref{MMF_WSR_cnst2}, \eqref{MMF_WSR_powCnst} \label{MMF_MSE11_cnst4}.
\end{align}
\end{subequations}

The problem defined in \eqref{MMF_MSE11_all} is hard to solve and there are no closed form expressions for the optimal precoders and PAC. Instead, in this section we propose an iterative precoder design algorithm. To do that, we need to split this problem into two different problems. In the first part of \eqref{MMF_MSE11_all}, we investigate the optimal precoders for a given set of PAC. Then, we update PAC using the updated precoders. For the first part we assume $\mathbf{A}$ is given and solve
\begin{subequations}\label{MMF_MSE2_all}
\begin{align}
\max_{\substack{\mathbf{P},\mathbf{R}, \\R_g,\mathbf{V}}} &\qquad R_g \label{MMF_MSE2} \\
 \text{s.t. }
&\qquad  R_{k,l} \leq  1- \xi_{k, i \rightarrow l}, \forall k \in \mathcal{K}, \forall l \in \mathcal{L}, \forall i \in \mathcal{I}, \label{MMF_MSE2_cnst1} \\
& \qquad \eqref{MMF_MSE11_cnst1}, \eqref{MMF_MSE11_cnst2},  \eqref{MMF_WSR_cnst2}, \eqref{MMF_WSR_powCnst}.
\end{align}
\end{subequations}where $\mathbf{V} = [V_{1, 1 \rightarrow 1},\ldots, V_{K, L \rightarrow L} ]$ and $\mathbf{b} = [b_{1, 1 \rightarrow 1},\ldots, b_{K, L \rightarrow L}]$ consist of all receivers and weights respectively. 
The optimization problem in \eqref{MMF_MSE2_all} is convex if either the precoder matrix $\mathbf{P}$ or the receiver matrix $\mathbf{V}$ is given. Thus, an iterative algorithm can solve \eqref{MMF_MSE2_all} sub-optimally, starting from an initial precoder $\mathbf{P}^{init}$. In each iteration, the algorithm can update $\mathbf{V}$ and $\mathbf{P}$ using a standard convex program solver such as CVX \cite{CVX}.

\subsection{Power Allocation for WMMSE}

In this subsection, we discuss the optimal PAC selection for the second part of \eqref{MMF_MSE11_all} for given precoders and receivers, and weights $b_{k, i \rightarrow l}$, which are already calculated using \eqref{b_weight}. 

Note that, PAC in each cluster are not related with the coefficients in other clusters, as inter-cluster power allocation is already a part of the precoder optimization in \eqref{MMF_MSE2_all}. Therefore, we can consider PAC optimization as intra-cluster power allocation and write a simplified problem for each cluster $k$ as
\begin{subequations}\label{MMF_PA_all}
\begin{align}
\max_{\substack{\mathbf{A}}} &\qquad R_{k,L} \label{MMF_PA} \\
 \text{s.t. }
&\qquad  R_{k,l}^{th} \leq  R_{k, i \rightarrow l}, \forall l \in \bar{\mathcal{L}}, \forall i \in \mathcal{I}, \label{MMF_PA_cnst1} \\
& \qquad \sum_{l=1}^{L} \alpha_{k,l}= 1,  \alpha_{k,l} \geq 0,  \; \forall l \in \mathcal{L} \label{MMF_PA_cnst2}.
\end{align}
\end{subequations}Although this problem is non-convex, we can make it affine using \eqref{rate_errrorCov1}. Then, \eqref{MMF_PA_all} becomes 
\begin{subequations}\label{MMF_PA2_all}
\begin{align}
\max_{\substack{\mathbf{A}}} &\qquad \log \left(  \alpha_{k,L} b_{k, L \rightarrow L} \right) \label{MMF_PA2} \\
 \text{s.t. }
&\qquad  \Psi_{k,l} \leq  \log \left(  \alpha_{k,l} b_{k, i \rightarrow l} \right), \forall l \in \bar{\mathcal{L}}, \forall i \in \mathcal{I}, \label{MMF_PA2_cnst1} \\
& \qquad \sum_{l=1}^{L} \alpha_{k,l}= 1,  \alpha_{k,l} \geq 0,  \; \forall l \in \mathcal{L} \label{MMF_PA2_cnst2}.
\end{align}
\end{subequations}Here, $\Psi_{k,l} = R_{k,l}^{th}$ \footnotemark\footnotetext{In the next subsection, we will alter this definition.}. The objective function in \eqref{MMF_PA2} is monotonically increasing in $\alpha_{k,L}$. As
\begin{align}
\alpha_{k,L} &= 1 - \sum_{l=1}^{L-1} \alpha_{k,l}, \label{alpha}
\end{align}we can restate \eqref{MMF_PA2_all} as
\begin{subequations}\label{MMF_PA3_all}
\begin{align}
\min_{\substack{\mathbf{A}}} &\qquad  \sum_{l=1}^{L-1} \alpha_{k,l} \label{MMF_PA3} \\
 \text{s.t. }
&\qquad  \frac{e^{\Psi_{k,l}}}{b_{k, i \rightarrow l}} \leq \alpha_{k,l}, \forall l \in \bar{\mathcal{L}}, \forall i \in \mathcal{I}, \label{MMF_PA3_cnst1} \\
& \qquad \alpha_{k,l} \geq 0,  \; \forall l \in \mathcal{L} \label{MMF_PA3_cnst2}.
\end{align}
\end{subequations}Then, we can obtain the optimal PAC as
\begin{align}
\alpha_{k,l}^{opt} &= \max_{i \in \mathcal{I}} \frac{e^{\Psi_{k,l}}}{b_{k, i \rightarrow l}}\label{optAlfa}, \forall l \in \bar{\mathcal{L}}.
\end{align}
and $\alpha_{k,L}^{opt}$ can be obtained using \eqref{alpha}. The optimal $\alpha_{k,l}^{opt}$ always satisfies \eqref{MMF_PA3_cnst2} $\forall l \in \bar{\mathcal{L}}$. However, this may not be true for the $L$-th user in each cluster $k$. Therefore, we update $\alpha_{k,l}^{(n)}$ as
\begin{align}\label{alfa_kL}
      \alpha_{k,l}^{(n)} =
      \begin{cases} 
      \alpha_{k,l}^{(n-1)}, &  \alpha_{k,L}^{opt}\leq 0 \\
      \alpha_{k,l}^{opt}, & \mathrm{otherwise} 
  \end{cases}, \quad \forall l \in \mathcal{L}.
\end{align}

\begin{algorithm}[!t]
    \caption{WMMSE1: MMF Algorithm with PA}
    \label{algorithm:algo_MMSECVX}
    \begin{algorithmic}[1]
    \State \textbf{Init:} $\mathbf{A}^{(0)}$, $E_{tx}$, $\Upsilon$, $\mathbf{P}^{(0)}$, $R_{k,l}^{th}$, $R_g^{(0)}=0$, $n_{max}$, $n=0$;\\
    \textbf{iterate} $\forall j, l, k$;
      \State \: $n = n +1$ 
      \State \: Compute ${V}_{k, i \rightarrow l}^{(n)}$ using \eqref{V_rec1} 
      \State \: Compute $\varepsilon_{k, i \rightarrow l}^{(n)}$ using \eqref{UMSE_k1}
      \State \: Compute $b_{k,i \rightarrow l}^{(n)}$ using \eqref{b_weight}
      \State \: \multiline{ Update $\left\{ \mathbf{P}_k^{(n)}, R_g^{(n)} \right \}$ by solving \eqref{MMF_MSE2_all} for given \\$V_{k, i \rightarrow l}^{(n)}$ and $b_{k, i \rightarrow l}^{(n)}$}
      \State \: Update $\mathbf{A}^{(n)}$ using \eqref{alfa_kL}
      \State \: $\mathbf{If}$ {$ (R_g^{(n)} - R_g^{(n-1)})/R_g^{(n-1)}  < \Upsilon $} $\textbf{or}$ $n=n_{max}$ $\mathbf{then}$
      \State \quad \quad Terminate 
      \State \: $\mathbf{else}$ $\mathbf{then}$ 
      \State \quad \quad Go to Step 2
    \end{algorithmic}
\end{algorithm} 

\subsection{A Low Complexity WMMSE Solution}
Algorithm \ref{algorithm:algo_MMSECVX} resorts to convex solvers in Step 7 to solve \eqref{MMF_MSE2_all} for given receivers and MMSE weights. Although, this results in the optimal solution in Step 7, it significantly increases computational complexity. In this subsection, we propose a low complexity solution. As the objective function and the constraints in \eqref{MMF_MSE2_all} are all continuously differentiable, we can make use of the KKT conditions to reduce the search space and thus to decrease complexity. 

When $\theta_k, \Gamma_{k,l}, \eta_{k, i \rightarrow l}$ and $\beta$ denote Lagrange multipliers, 
the Lagrangian objective function of \eqref{MMF_MSE2_all} is written as
\begin{align}
&h(\mathbf{P}, \mathbf{R}, R_g, \mathbf{V} ) = - R_g + \sum_{k=1}^{K} \theta_k( R_g -  R_{k,L} )\nonumber \\
&+ \sum_{k=1}^{K}\sum_{l=1}^{L}\sum_{i=l}^{L}\eta_{k, i \rightarrow l}  (R_{k,l} - 1 + \xi_{k,  \rightarrow l} ) \nonumber \\
&+ \sum_{k=1}^{K}\sum_{l=1}^{L-1} \Gamma_{k,l} (R_{k,l}^{th} - R_{k,l}) + \beta ( \Tr (\mathbf{P}\mathbf{P}^H) - E_{tx}).\label{lagrangian_h}
 \end{align}The optimal precoders and the receivers have to satify the KKT conditions for \eqref{lagrangian_h}, and are given in the following theorem. 

\begin{theorem}\label{theorem_h1}
For the optimization problem defined in \eqref{MMF_MSE2_all}, the following receivers ${V}_{k, i \rightarrow l}$, the Lagrange multiplier $\beta$, and the transmit precoder vectors $\mathbf{p}_k$ satisfy the KKT conditions.
\begin{align}
{V}_{k, i \rightarrow l} &= \alpha_{k,l}\mathbf{p}_k^H \mathbf{h}_{k,i} T_{k, i \rightarrow l}^{-1},\label{V_rec1_theo}\\
\beta &= \frac{1}{E_{tx}}\left[\sum_{k=1}^{K} \sum_{l=1}^{L}\sum_{i=l}^{L}\eta_{k, i \rightarrow l} b_{k, i \rightarrow l} |{V}_{k, i \rightarrow l}|^2 \right], \label{beta} \\
 \mathbf{p}_k &= \bigg[ \beta\mathbf{I} + \sum_{l=1}^{L} \sum_{i=l}^{L}\sum_{j=l}^{L} \alpha_{k,j} \eta_{k, i \rightarrow l} b_{k, i \rightarrow l}  \mathbf{h}_{k,i} |{V}_{k, i \rightarrow l}|^2 \mathbf{h}_{k,i}^{H}  \nonumber \\
 &\quad + \sum_{t=1, t\neq k}^{L}  \sum_{l=1}^{L}  \sum_{i=l}^{L}\eta_{t,i \rightarrow l} b_{t, i \rightarrow l}
\mathbf{h}_{t,i} |{V}_{t,i \rightarrow l}|^2  \mathbf{h}_{t,i}^{H} \bigg]^{-1}\nonumber\\
&\:\quad \times \left[ \sum_{l=1}^{L} \sum_{i=l}^{L}\eta_{k, i \rightarrow l} b_{k, i \rightarrow l}\alpha_{k,l}  \mathbf{h}_{k,i} {V}_{k, i \rightarrow l}^{^\ast}  \right].
\label{Pk_theo}
\end{align}
\end{theorem}
\begin{IEEEproof}
The proof is provided in Appendix \ref{derive_Pk}.
\end{IEEEproof}
\begin{remark}
The receiver ${V}_{k, i \rightarrow l}$ in (\ref{V_rec1_theo}) is exactly equal to the MMSE receiver ${V}_{k, i \rightarrow l}^{mmse}$ given in (\ref{V_rec1}).
\end{remark}
\begin{remark}
When the optimal MMSE receiver ${V}_{k, i \rightarrow l}^{mmse}$ and the weights $b_{k,i\rightarrow l}^{mmse}$ in \eqref{b_weight} are substituted in $\xi_{k,i \rightarrow l}$ of \eqref{ksi_u}, then $\xi_{k,i \rightarrow l}$ becomes equal to $\xi_{k,i \rightarrow l}^{mmse}$. 
\end{remark}

Utilizing Theorem \ref{theorem_h1}, we propose solving for the receivers \eqref{V_rec1_theo}, the Lagrange multiplier \eqref{beta} and the precoders \eqref{Pk_theo} in an iterative fashion in Algorithm \ref{algorithm:algo_WMMSECForm}. However, calculating the Lagrange multipliers set $\{\theta_k, \Gamma_{k,l}, \eta_{k, i \rightarrow l}, \allowbreak \forall i, l, k\}$ 
is not trivial. In \cite{Li2003_ExponentialPenaltyMethod}, an exponential penalty method is suggested to solve min-max type problems. According to the exponential penalty method, in each iteration of the algorithm, we update $\{\theta_k, \Gamma_{k,l}, \eta_{k, i \rightarrow l}\}$ as
\begin{align}
\theta_k &= \frac{\exp\left\{\nu \left( R_g - R_{k,L} \right)\right\}}{ \sum_{k=1}^{K} \exp\left\{\nu \left( R_g - R_{k,L}\right)\right\}}, \forall k \in \mathcal{K}, \label{theta_k}\\
 \Gamma_{k,l}  &= \exp\left\{\nu \left( R_{k,l}^{th} - R_{k,l} \right)\right\}, \forall k \in \mathcal{K}, \forall l \in \bar{\mathcal{L}}, \label{gamma_lk}\\
\eta_{k, i \rightarrow l} &=  \Gamma_{k,l} \frac{\exp\left\{\nu \left(R_{k,l} - R_{k, i \rightarrow l} \right)\right\}}{\sum_{i=l}^{L}  \exp\left\{\nu \left(R_{k,l} - R_{k, i \rightarrow l}\right)\right\}}, \nonumber \\
\eta_{k, L \rightarrow L} &= \theta_k, \forall k \in \mathcal{K}, \forall l \in \bar{\mathcal{L}} , \forall i \in \mathcal{I}.\label{eta_jlk}
\end{align}In the above equations, $\nu$ is a constant and as long as $\nu \geq (\log KL)/\epsilon $, the solution is $\epsilon$-optimal. Note that, this choice satisfies the KKT conditions on $\{\theta_k, \Gamma_{k,l},  , \eta_{k, i \rightarrow l}\}$ since $\sum_{k=1}^{K} \theta_k = 1$, $\sum_{i=l}^{L}  \eta_{k, i \rightarrow l} = \Gamma_{k,l}, \forall l \in \bar{\mathcal{L}}$ and $\eta_{k, L \rightarrow L} = \theta_k$, $\theta_k \geq 0, \Gamma_{k,l} \geq 0, \eta_{k, i \rightarrow l} \geq 0$. 

\begin{algorithm}[!t]
    \caption{WMMSE2: Low Complexity MMF Algorithm with PA}
    \label{algorithm:algo_WMMSECForm}
    \begin{algorithmic}[1]
    \State \multiline{\textbf{Init:} $\epsilon$, $\mathbf{A}^{(0)}$, $E_{tx}$, $\Upsilon$, $\Delta$, $\mathbf{P}^{(0)}$, $R_{k,l}^{th}$, $R_g^{(0)}=0$, \\$\nu = \log (KL)/\epsilon $, $n_{max}$, $n=0$;}\\
    \textbf{iterate} $\forall j, l, k$;
      \State \: $n = n +1$ 
      \State \: Compute ${V}_{k, i \rightarrow l}^{(n)}$ using \eqref{V_rec1_theo}
      \State \: Compute $\varepsilon_{k, i \rightarrow l}^{(n)}$ using \eqref{UMSE_k1}
      \State \: Compute $b_{k,i \rightarrow l}^{(n)}$ using \eqref{b_weight}
      \State \: Compute $\Gamma_{k,l}^{(n)}$ using \eqref{gamma_lk}
      \State \: Compute $\theta_k^{(n)}$ using \eqref{theta_k}
      \State \: Compute $\eta_{k,i \rightarrow l}^{(n)}$ using \eqref{eta_jlk}
      \State \: Compute $\beta^{(n)}$ using \eqref{beta}
      \State \: Compute $\mathbf{P}^{(n)}$ using \eqref{Pk_theo}
      \State \: Scale $\mathbf{P}^{(n)}$ such that  $\Tr (\mathbf{P}^{(n)} {\mathbf{P}}^{(n)^H}) = E_{tx} $  
      \State \: Update $\mathbf{A}^{(n)}$ using \eqref{alfa_kL}
      \State \: $\mathbf{If}$ {$ (R_g^{(n)} - R_g^{(n-1)})/R_g^{(n-1)}  < \Upsilon $} $\textbf{or}$ $n=n_{max}$ $\mathbf{then}$
      \State \quad \: $\mathbf{If}$ \eqref{MMF_MSE1_cnst2} satisfied $\mathbf{then}$ 
      \State \quad \quad \: Terminate 
      \State \quad \:  $\mathbf{else}$ $\mathbf{then}$ \State \quad \quad \: $\nu = \nu + \Delta$, Go to Step 2
      \State \: $\mathbf{else}$ $\mathbf{then}$ 
      \State \quad \quad \: Go to step 2
    \end{algorithmic}
\end{algorithm} 

In each iteration, Algorithm \ref{algorithm:algo_WMMSECForm} increases the objective function, since there is a total power constraint. Thus, the proposed WMMSE algorithm converges to an upper limit. This limit is within an $\epsilon$ neighborhood of a local optimum, as the algorithm utilizes the equations found via the KKT conditions, and the exponential penalty method is employed. Following similar steps as in \cite[Section IV-A]{Christensen2008_WeightedSumRate} and  \cite{Kaleva2016_DecentralizedSumRateMaximization}, one can prove convergence in full detail. 

\subsubsection{Power Allocation for Low Complexity WMMSE}
Algorithm \ref{algorithm:algo_MMSECVX} always returns a solution at Step 7, as the CVX approach returns the final result for \eqref{MMF_MSE2_all} for given receivers and weights. On the other hand, the low-complexity WMMSE approach may not be feasible in each iteration, as it only provides a step in the favorable direction in each iteration. Therefore, in Algorithm \ref{algorithm:algo_WMMSECForm} at Step 12, the updated precoder $\mathbf{P}^{(n)}$ may not satisfy the threshold rate constraints in \eqref{MMF_MSE1_cnst2}, and the algorithm may not find a feasible PAC at Step 13. One approach would be to skip power optimization, immediately update $\nu$ and proceed with the next iteration. However, we choose to find the best PAC that satisfies the current achievable rates. Thus, we update $\Psi_{k,l}$ in \eqref{MMF_PA2_cnst1} in each iteration as 
\begin{align}
\Psi_{k,l}^{(n)} &= \min \left( R_{k,l}^{th}, R_{k,l}^{(n)}\right).
\end{align}



\section{Numerical Results}\label{sec:sim}

In this section, we present numerical results to evaluate the performance of the proposed transmission strategies given in Algorithms \ref{algorithm:algo_SCA}, \ref{algorithm:algo_MMSECVX} and \ref{algorithm:algo_WMMSECForm}. We compare these algorithms with OMA, MULP and RS in terms of MMF rates, energy efficiency and computational complexity. All three algorithms we propose carry out power optimization. We also compare them with their fixed power allocation versions.

\subsection{Orthogonal Multiple Access, Multiuser Linear Precoding and Rate Splitting}
Before presenting any simulation results, in this subsection, we first describe the schemes used as benchmarks: OMA, MULP and RS. 

\subsubsection{OMA}
In OMA, the transmission time is divided into $L$ equal slots. The base station communicates with the $l$-th strongest users in each cluster in each time slot-$l$. The input data vector for time slot $l$ is denoted as $\mathbf{s}^{l,OMA} = {[s_{1,l},\ldots,s_{K,l}]}^T$ $\in \mathbb{C}^{K \times 1}$. We assume all $s_{k,l}$  are independent and
$\mathbb{E}\{{s}_{k,l}{s}_{k,l}^{\ast}\} =1$. The input data vector $\mathbf{s}^{l,OMA}$ is linearly processed by a precoder matrix $\mathbf{P}^{l,OMA} = [\mathbf{p}_1^{l,OMA},\ldots,\mathbf{p}_{K}^{l,OMA}]$  $\in \mathbb{C}^{M \times K}$, where the precoding vector $\mathbf{p}_k^{l,OMA} \in \mathbb{C}^{M\times 1}$ is dedicated to the $k$-th user in time slot-$l$. The overall transmit data vector $\mathbf{x}^{l,OMA}$ $\in \mathbb{C}^{M\times 1}$ at the base station can be written as $\mathbf{x}^{l,OMA} = \mathbf{P}^{l,OMA}\mathbf{s}^{l,OMA}$. Then, the SINR at user-$k$ in time slot-$l$ is given by 
\begin{align}
\gamma_{k}^{l,OMA} &= \frac{\left|\mathbf{h}_{k,l}^{H} \mathbf{p}_k^{l,OMA}\right|^2}{\sum_{i=1, i\neq k}^{K}  \left|\mathbf{h}_{k,l}^{H} \mathbf{p}_i^{l,OMA}\right|^2  + 1} \label{SINR_OMA}
\end{align}and the corresponding rate expression is calculated as $R_{k,l}^{OMA} = \frac{1}{L}\log(1 + \gamma_{k}^{l,OMA})$. 

Given these assumptions, the MMF OMA problem is equivalent to providing fairness in the last time slot, while satisfying the threshold rate constraints in earlier time slots. We can formulize the MMF OMA optimization problem as 
\begin{subequations}\label{MMF_OMA_all}
\begin{align}
\max_{\mathbf{P}^{l,OMA}, \forall l \in \mathcal{L}} &\; \min_{k \in \mathcal{K}} R_{k,L}^{OMA}\label{MMF_OMA} \\
 \text{s.t. }&\; R_{k,l}^{th} \leq  R_{k,l}^{OMA},\forall k \in \mathcal{K}, \forall l \in \bar{\mathcal{L}} \label{MMF_OMA_cnst2}\\
 &\;  \Tr (\mathbf{P}^{l,OMA}\mathbf{P}^{{l,OMA}^H}) \leq E_{tx}, \forall l \in \mathcal{L}. \label{MMF_OMA_powCnst}
\end{align}\end{subequations} 
Then, the MMF OMA rate $R^{OMA}$ can be calculated as \[R^{OMA} = \min_{k \in \mathcal{K}} {R_{k,L}^{OMA}}\] using the optimal precoders ${\mathbf{P}^{l,OMA}}^*$, $\forall l \in \mathcal{L}$ that solve \eqref{MMF_OMA_all}. Note that precoders ${\mathbf{P}^{l,OMA}}^*$, $\forall l \in \bar{\mathcal{L}}$ are required to satisfy the rate constraints in \eqref{MMF_OMA_cnst2}, whereas ${\mathbf{P}^{L,OMA}}^*$ provides fairness among the strongest users in each cluster.


\subsubsection{MULP} In MULP precoding, the base station transmits data to all $KL$ users simultaneously. The input data vector is denoted as  $\mathbf{s}^{MULP} = {[s_{1,1},\ldots,s_{1,L}, \ldots, s_{K,1}, \ldots, s_{K,L} ]}^T$ $\in \mathbb{C}^{KL \times 1}$. We assume all $s_{k,l}$  are independent and
$\mathbb{E}\{{s}_{k,l}{s}_{k,l}^{\ast}\} =1$. The input data vector $\mathbf{s}^{MULP}$ is linearly processed by a precoder matrix $\mathbf{P}^{MULP} = [\mathbf{p}_{1,1}^{MULP},\ldots,\mathbf{p}_{1,L}^{MULP}, \ldots, \mathbf{p}_{K,1}^{MULP},\ldots,\mathbf{p}_{K,L}^{MULP}]$  $\in \mathbb{C}^{M \times KL}$, where the precoding vector $\mathbf{p}_{k,l}^{MULP} \in \mathbb{C}^{M\times 1}$ is dedicated to the $l$-th user in the $k$-th cluster. Then, the overall transmit data vector $\mathbf{x}^{MULP}$ $\in \mathbb{C}^{M\times 1}$ at the base station can be written as $\mathbf{x}^{MULP} = \mathbf{P}^{MULP}\mathbf{s}^{MULP}$. The SINR at user-$l$ in the $k$-th cluster is given by
\begin{align}
&\gamma_{k,l}^{MULP}= \nonumber \\
& \frac{|\mathbf{h}_{k,l}^{H} \mathbf{p}_{k,l}^{MULP}|^2}{\sum_{\substack{j=1\\j\neq l}}^{L}  |\mathbf{h}_{k,l}^{H} \mathbf{p}_{k,j}^{MULP}|^2 + \sum_{\substack{i=1\\ i\neq k}}^{K} \sum_{l=1}^{L} |\mathbf{h}_{i,l}^{H} \mathbf{p}_{i,l}^{MULP}|^2  + 1}, \label{SINR_MULP}
\end{align}and the corresponding rate expression is calculated as $R_{k,l}^{MULP} = \log(1 + \gamma_{k, l}^{MULP})$. 

For a fair comparison, we assume that fairness among the strongest users is needed while satisfying the threshold rate constraints on other users. The MMF MULP problem is written as 
\begin{subequations}\label{MMF_MULP_all}
\begin{align}
\max_{\mathbf{P}^{MULP}} &\; \min_{k \in \mathcal{K}} R_{k,L}^{MULP}\label{MMF_MULP} \\
 \text{s.t. }&\;R_{k,l}^{th} \leq  R_{k,l}^{MULP}, \forall k \in \mathcal{K}, \forall l \in \bar{\mathcal{L}}\label{MMF_MULP_cnst2}\\
 &\;  \Tr (\mathbf{P}^{MULP}\mathbf{P}^{MULP^H}) \leq E_{tx}. \label{MMF_MULP_powCnst}
\end{align}
\end{subequations} 
Then, the MMF MULP rate $R^{MULP}$ can be calculated as 
\[R^{MULP} = \min_{k \in \mathcal{K}} {R_{k,L}^{MULP}}\] using the optimal precoder ${\mathbf{P}^{MULP}}^*$ that solve \eqref{MMF_MULP_all}.

\subsubsection{1-Layer RS}

In 1-Layer RS, we use the same signal model proposed in \cite{Mao2018}. In this strategy, the message stream of the $l$-th user in the $k$-th cluster is split into {common} and {private} parts. The common part is at rate $C_{k,l}^{RS}$ and the private part is at rate $R_{k,l}^{RS}$. The common parts are collectively encoded as a common message $s_{c}$ at rate $R_{c}^{RS}= \sum_{k\in\mathcal{K}}\sum_{l\in\mathcal{L}}C_{k,l}^{RS}$. The private messages are encoded as $s_{k,l,p}$. To send all the messages, the base station encodes the input data vector $\mathbf{s}
^{RS}= {[{s}_c, s_{1,1,p},\ldots,s_{1,L,p},\ldots, {s}_{K,1,p} \ldots,{s}_{K,L,p}]}^T$ $\in \mathbb{C}^{(KL+1) \times 1}$ by a precoder matrix $\mathbf{P}^{RS} = [\mathbf{p}_c^{RS}, \mathbf{p}_{k,1}^{RS}, \ldots, \mathbf{p}_{k,L}^{RS}, \ldots, \mathbf{p}_{K,l}^{RS}, \ldots, \mathbf{p}_{K,L}^{RS}]$. Here, $\mathbf{p}_c^{RS}$ and  $\mathbf{p}_{k,l}^{RS} \in \mathbb{C}^{M\times 1}$ respectively indicate the precoder vectors for the common data ${s}_c$ and the private data $s_{k,l,p}$. The base station transmits $\mathbf{x}
^{RS}$ $\in \mathbb{C}^{M\times 1}$, which is equal to $\mathbf{x}^{RS} = \mathbf{P}^{RS}\mathbf{s}^{RS}$
Then, the SINR at the $l$-th user in the $k$-th cluster for common and private data messages respectively become 
\begin{align}
&\gamma_{k,l,c}^{RS} = \frac{|\mathbf{h}_{k,l}^{H} \mathbf{p}_{c}^{RS}|^2}{ \sum_{\substack{i=1}}^{K} \sum_{l=1}^{L} |\mathbf{h}_{k,l}^{H} \mathbf{p}_{k,l}^{RS}|^2  + 1}, \label{SINR_RS_c}\\
&\gamma_{k,l,p}^{RS}= \nonumber \\
& \frac{|\mathbf{h}_{k,l}^{H} \mathbf{p}_{k,l}^{RS}|^2}{\sum_{\substack{j=1\\j\neq l}}^{L}  |\mathbf{h}_{k,l}^{H} \mathbf{p}_{k,j}^{RS}|^2 + \sum_{\substack{i=1\\ i\neq k}}^{K} \sum_{l=1}^{L} |\mathbf{h}_{i,l}^{H} \mathbf{p}_{i,l}^{RS}|^2  + 1},
\end{align}and the corresponding rate expressions are calculated as $R_{k,l,c}^{RS} = \log(1 + \gamma_{k, l,c}^{RS})$ and $R_{k,l,p}^{RS} = \log(1 + \gamma_{k,l,p}^{RS})$. 
As the common rate has to be decoded by all users, we define $R_{c}^{RS} = \min_{k\in\mathcal{K}} \min_{l\in\mathcal{L}} R_{k,l,c}^{RS}$. Then, the MMF RS problem can be stated as 
\begin{subequations}\label{MMF_RS_all}
\begin{align}
\max_{\mathbf{P}^{RS}} &\quad \min_{k \in \mathcal{K}} \left(C_{k,L}^{RS} + R_{k,L}^{RS}\right)\label{MMF_RS} \\
 \text{s.t. }&\quad R_{k,l}^{th} \leq  \left(C_{k,l}^{RS} + R_{k,l}^{RS}\right), \forall k \in \mathcal{K}, \forall l \in \bar{\mathcal{L}}\label{MMF_RS_cnst2}\\
 & \quad \sum_{k\in\mathcal{K}}\sum_{l\in\mathcal{L}}C_{k,l}^{RS} \leq R_{c}^{RS},\\
  & \quad 0 \leq C_{k,l}^{RS}, \forall k \in \mathcal{K}, \forall l \in \mathcal{L}  \\
 &\quad  \Tr (\mathbf{P}^{RS}\mathbf{P}^{RS^H}) \leq E_{tx}. \label{MMF_RS_powCnst}
\end{align}
\end{subequations}
As a result, the MMF RS rate $R^{RS}$ becomes
\[R^{RS} = \min_{k, \in \mathcal{K} } \left({C_{k,L}^{RS}} + {R_{k,L}^{RS}}\right)\] employing the optimal precoder that solves \eqref{MMF_RS_all}.

To solve all optimization problems stated for OMA, MULP and RS, we first find their equivalent weighted MMSE problems and solve them in an iterative fashion as done in Algorithm \ref{algorithm:algo_MMSECVX} in Section \ref{sec:WMMSE}.

\subsection{Assumptions}
In the simulations, the entries in $\mathbf{\tilde{h}}_{k,l}$ are assumed to be i.i.d. circularly symmetric complex Gaussian random variables with zero mean and unit variance. 
The path loss exponent is set to $\rho= 4$. 
The users are  uniformly distributed in a circular region of radius 1. These users are clustered according to the scheme proposed in \cite[Algorithm 1, Figure 3]{Ali2016_DynamicUser}.  
In this clustering scheme, the aim is to put users, which have highly different effective channel gain magnitudes in the same cluster. 
For example, for $L= 2$, the base station puts the user with the highest effective channel gain magnitude in the same cluster with the worst user among all users. The second best and and the second worst users are grouped as a second cluster. The remaining clusters are formed in a similar fashion. Note that for all the NOMA schemes, the base station has to inform the users about their order and the other users in their own cluster so that users within a cluster can perform SIC. 

\begin{figure}[!t]
\vspace{-0.0in}
\centering
\hspace*{-0.0in}
\includegraphics[width= 3.7in]{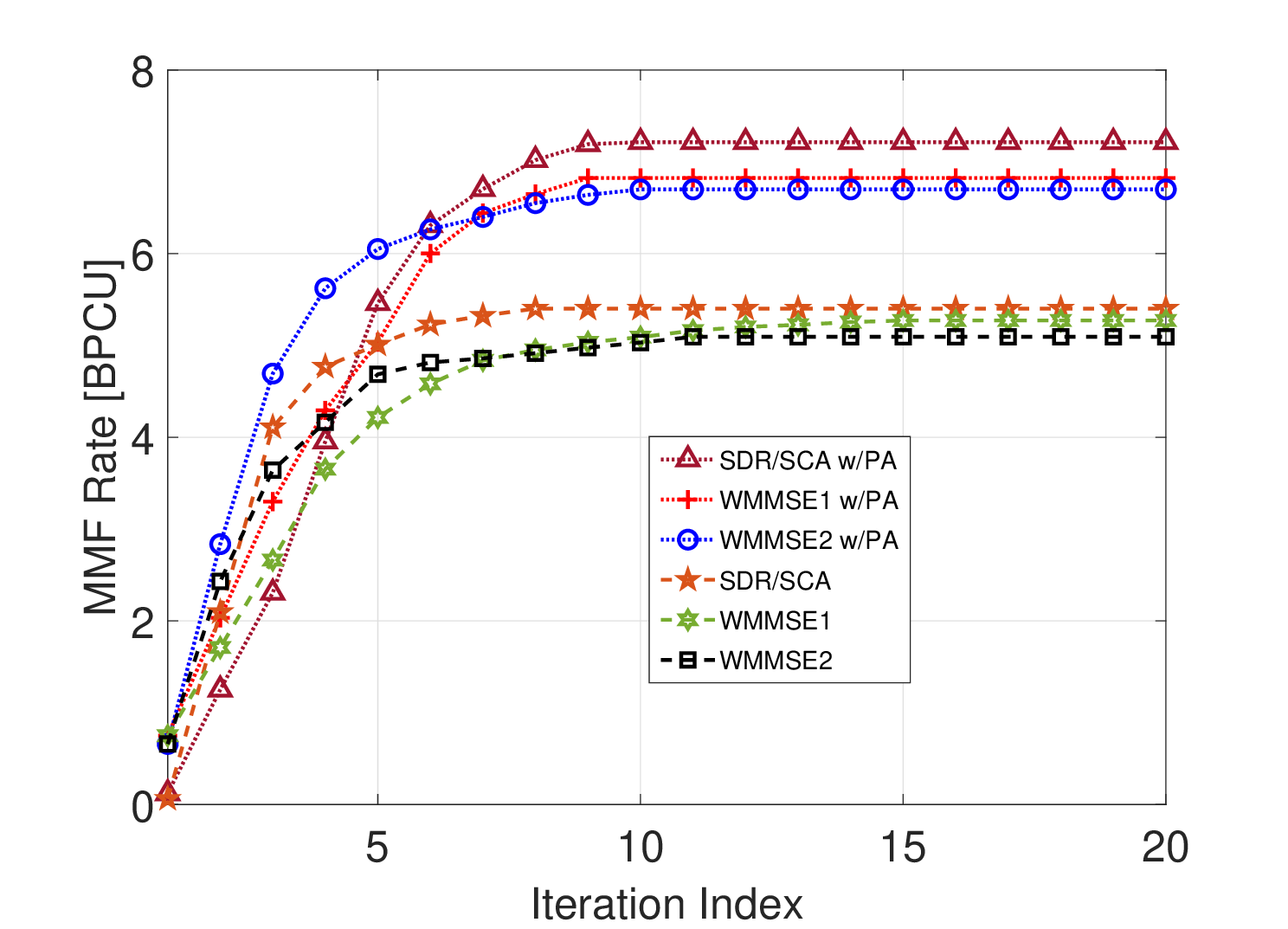}
\vspace{-0.10in}
\caption{MMF rate convergence performance of proposed algorithms for $M=K=3, L=2$, $R_{k,l}^{th} = 0.1$ bpcu. Transmit SNR is set to 15 dB. }\label{fig:convergenge_332}
\vspace{-0.00in}
\end{figure}

For the fixed power allocation versions of Algorithms \ref{algorithm:algo_SCA}, \ref{algorithm:algo_MMSECVX} and \ref{algorithm:algo_WMMSECForm}, we assume the power allocation scheme suggested in \cite[Table 1]{Ali2016_DynamicUser}, which assigns more power to weak users and less power to strong users. This idea is in line with power domain NOMA and widely used in the literature \cite{XSun2015_NOMAWithWSR},  \cite{Ali2017_NOMAForDownlinkMUMIMOSystems}. This fixed power allocation vector is also used as the initial value of $\mathbf{A}^{(0)}$ in Algorithms \ref{algorithm:algo_SCA}, \ref{algorithm:algo_MMSECVX} and \ref{algorithm:algo_WMMSECForm}.
 
For Algorithms \ref{algorithm:algo_SCA}, \ref{algorithm:algo_MMSECVX} and \ref{algorithm:algo_WMMSECForm}, the presented results are averaged over $10^2$ channel realizations. The maximum number of iterations $n_{max}$ is limited to $100$ and $\Upsilon$ are set to $10^{-3}$. The transmit signal to noise ratio (SNR) is defined as $E_{tx}/\sigma^2$. Here $\sigma^2$ is the noise variance and set to~$1$. For Algorithm~\ref{algorithm:algo_WMMSECForm},  $\epsilon$ and $\Delta$ are set to $10^{-3}$ and $3$ respectively. The parameter $\Delta$ is used to tune the algorithm to satisfy the rate constraint $R_{k,l}^{th}$. Finally, if a particular algorithm is infeasible, we set its MMF rate to zero to make a fair comparison among all algorithms under consideration \cite{XSun2018_JointBeamformingandPA}.

In the following simulation results, we consider algorithm convergence, MMF rate and energy efficiency results for different settings. Rates are expressed in terms of bits per channel use (bpcu).

\subsection{Simulation Results}

Fig. \ref{fig:convergenge_332} shows the convergence behavior of the proposed schemes given by Algorithms~\ref{algorithm:algo_SCA}, \ref{algorithm:algo_MMSECVX} and \ref{algorithm:algo_WMMSECForm} with and without PA for $M=3, K=3, L=2$, $R_{k,l}^{th} = 0.1$ $\forall k \in \mathcal{K}$, $\forall l \in \bar{\mathcal{L}}$, when the total transmit power is set to $15$ dB. The initial precoder matrix, $\mathbf{P}^{(0)}$ in Algorithms \ref{algorithm:algo_SCA}, \ref{algorithm:algo_MMSECVX} and \ref{algorithm:algo_WMMSECForm}
is assumed to be the identity matrix, scaled to satisfy the power constraint. The figure confirms that the proposed algorithms converge fast. 

\begin{figure}[!t]
\vspace{-0.0in}
\centering
\hspace*{-0.0in}
\includegraphics[width=3.7in]{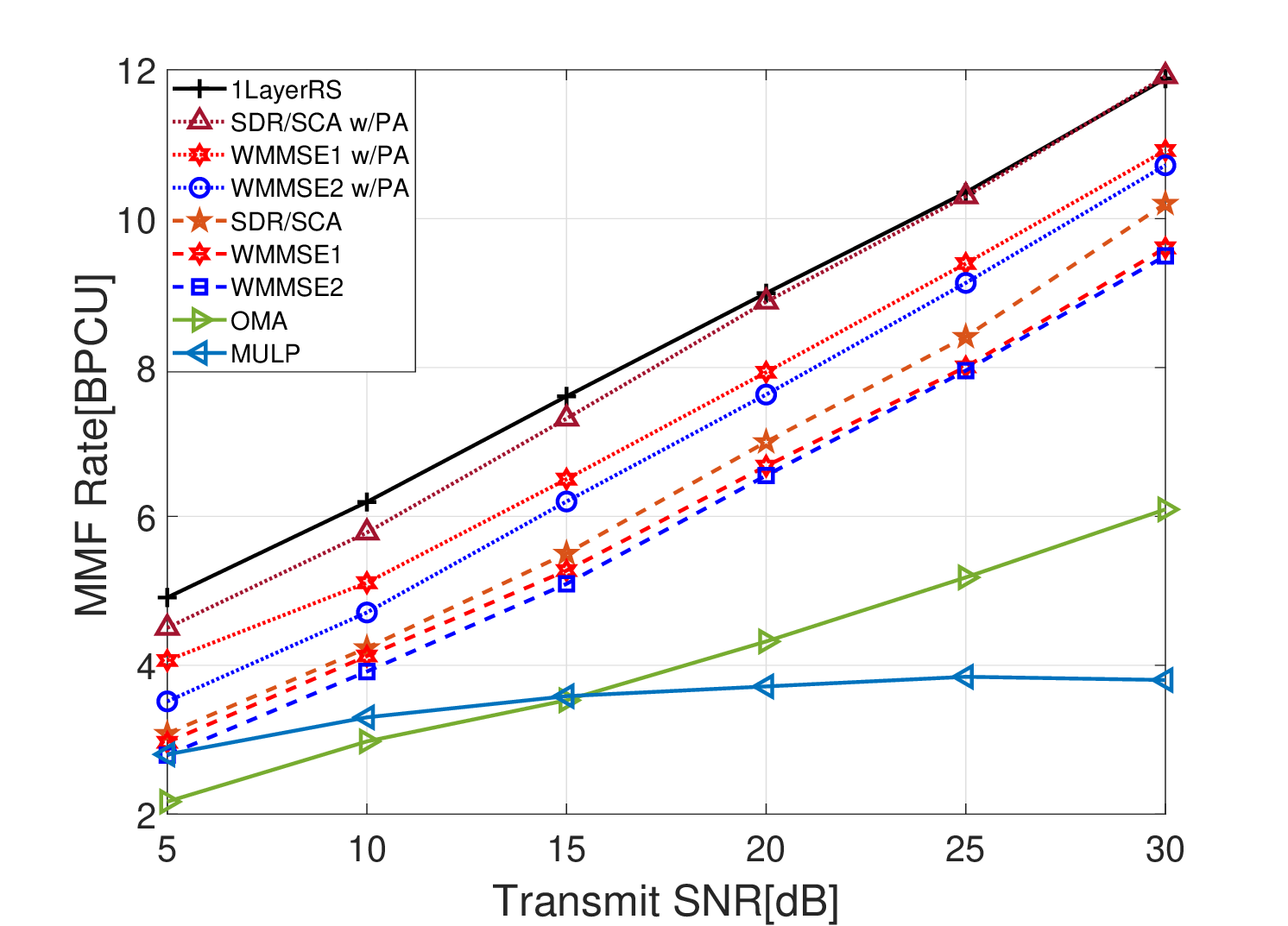}
\vspace{-0.10in}
\caption{MMF rates for different precoder schemes for $M=K=3, L=2$, $R_{k,l}^{th} = 0.1$ bpcu.}\label{fig:MMF_332}
\vspace{-0.00in}
\end{figure}

\begin{figure}[t]
\vspace{-0.0in}
\centering
\hspace*{-0.0in}
\includegraphics[width=3.7in]{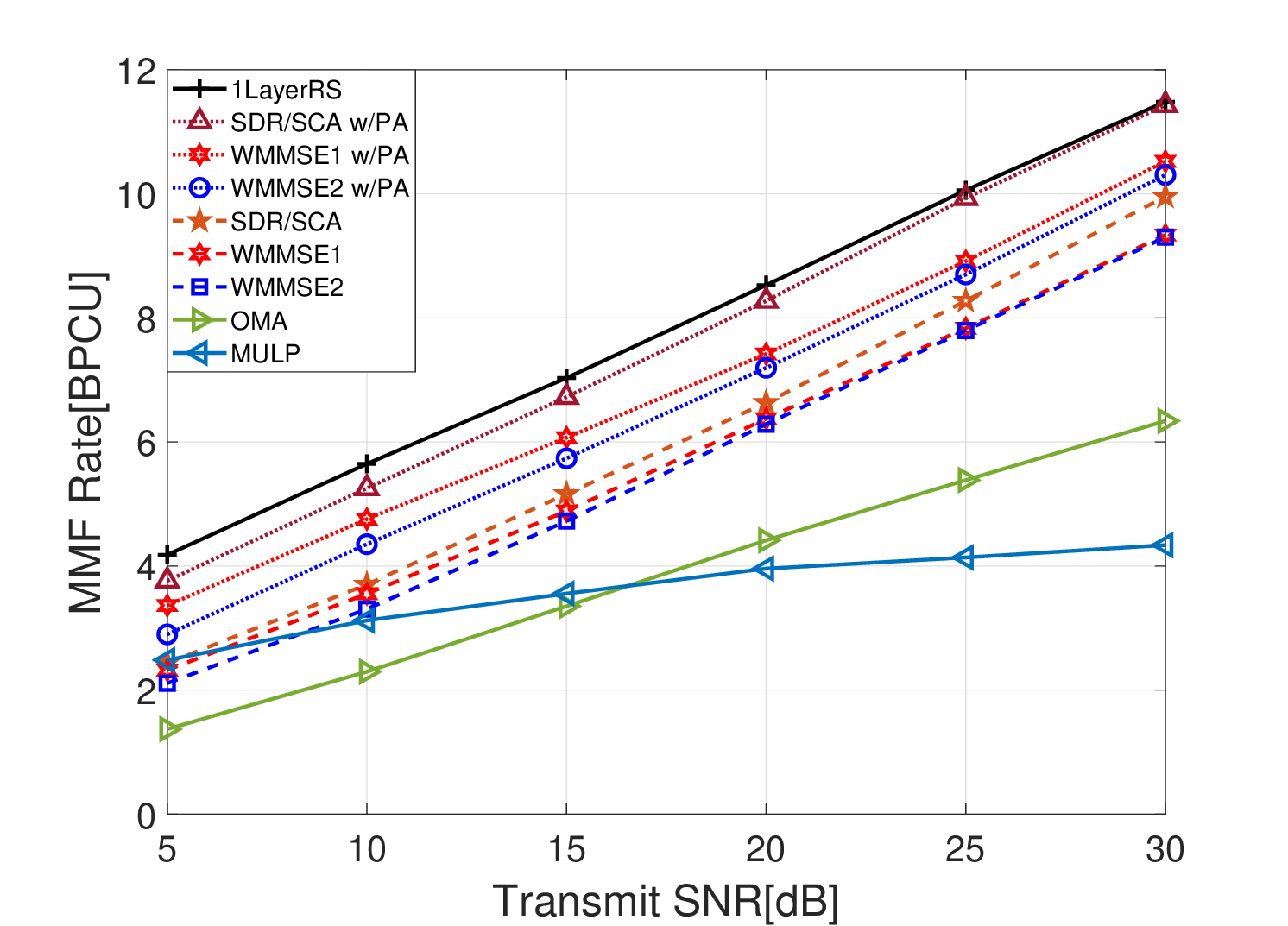}
\vspace{-0.1in}
\caption{MMF rates for different precoder schemes for $M=K=4, L=2$, $R_{k,l}^{th} = 0.1$ bpcu.}\label{fig:MMF_442}
\vspace{-0.00in}
\end{figure}
Figs. \ref{fig:MMF_332} and \ref{fig:MMF_442} compare MMF rates for the proposed algorithms with 1-layer RS, OMA and MULP schemes for $R_{k,l}^{th} = 0.1$ bits $\forall k \in \mathcal{K}, \forall l \in \bar{\mathcal{L}}$ for $M=3, K=3, L=2$, and for $M=4, K=4, L=2$ respectively. We observe that 1-layer RS has the best performance in terms of MMF rates. It can effectively mitigate interference by adjusting the common message rate. Algorithm \ref{algorithm:algo_SCA} (SDR/SCA w/PA) has similar performance with 1-layer RS and as SNR increases the gap between the two algorithms diminish. Both Algorithms \ref{algorithm:algo_MMSECVX} (WMMSE1 w/PA) and \ref{algorithm:algo_WMMSECForm} (WMMSE2 w/PA) perform worse than Algorithm \ref{algorithm:algo_SCA}. This is because semi-definite programming with successive convex approximation is an effective approximation. In each iteration, the constraints in \eqref{MMF_WSR5_cnst1} and \eqref{MMF_WSR5_cnst2} become tighter and \eqref{MMF_WSR5_all} approaches the original optimization problem in \eqref{MMF_WSR1_all}. 
As expected, Algorithms \ref{algorithm:algo_MMSECVX} and \ref{algorithm:algo_WMMSECForm} have similar results. All algorithms are several dB better than their fixed power allocation versions (SDR/SCA, WMMSE1, WMMSE2). MULP is very inefficient in interference management, and displays very poor performance. The MMF rate for MULP converges for high SNR.

From Figs. \ref{fig:MMF_332} and \ref{fig:MMF_442} we also observe that all Algorithms \ref{algorithm:algo_SCA}, \ref{algorithm:algo_MMSECVX} and \ref{algorithm:algo_WMMSECForm} (with or without power allocation)
and the RS scheme present full degrees of freedom (DoF); i.e. 1. DoF is calculated as the MMF rate (in bpcu) over $\log_2 \mathrm{SNR}$ \cite{Joudeh2017_RateSplittingforMaxMinFair}. While OMA can accommodate all users in each time slot, it suffers from time division and its DoF is limited with 0.5. Although a detailed DoF analysis is out of the scope of this paper, we conjecture that the DoF for MULP for the overloaded settings in Figs. \ref{fig:MMF_332} and \ref{fig:MMF_442} is 0. This is because, the MMF rate calculation for MULP is similar to the MMF rate calculation for the designated beamforming scheme in the multigroup multicasting scenario examined in \cite{Joudeh2017_RateSplittingforMaxMinFair}. For the latter, the DoF is proved to be 0 either for $M=3$ and there are 3 groups with 2 users each or for $M=4$ and there are 4 groups with 2 users each.

\begin{table}[t]
\centering
\begin{tabular}{|c|c|c|}
  \hline
   \multirow{1}{*}{} & \multicolumn{2}{|c|}{ RUN TIME IN SECONDS}  
    \\ \hline
  \multirow{1}{*}{ALGORITHM} & $M=K=3, L=2$ &  $M=K=4, L=2$
    \\ \hline
 \multirow{1}{*}{WMMSE2} & 6.38& 13.42\\ \hline
  \multirow{1}{*}{WMMSE2 w/PA} & 38.35 & 51.52\\ \hline
  \multirow{1}{*}{SDR/SCA} & 442.13 & 591.21 \\ \hline
  \multirow{1}{*}{SDR/SCA w/PA} & 491.23 & 677.29 \\ \hline
 \multirow{1}{*}{WMMSE1} & 2134.45 & 3743.80 \\ \hline
  \multirow{1}{*}{WMMSE1 w/PA} & 2752.55 & 4167.50 \\ \hline
 \multirow{1}{*}{1-layer RS} & 60457.80 & 75994.30 \\ \hline
  \end{tabular}
  \vspace{0.07in}
 \caption{Complexity of Algorithms}\label{table_1}
 \vspace{-0.0in}
\end{table}

Figs. \ref{fig:MMF_332} and \ref{fig:MMF_442} should be interpreted together with the complexity results given in Table \ref{table_1}. Table \ref{table_1} shows the complexity of all the algorithms under consideration. We observe that Algorithm \ref{algorithm:algo_WMMSECForm} has the least complexity either with or without power optimization. For Algorithms \ref{algorithm:algo_SCA}, \ref{algorithm:algo_MMSECVX} and \ref{algorithm:algo_WMMSECForm}, power optimization does not change algorithm complexity and run time values are on the same order. Although 1-layer RS has the highest MMF rates in Figs.~\ref{fig:MMF_332} and \ref{fig:MMF_442}, it also has the highest complexity. The run time for 1-layer RS is 3-4 orders of magnitude larger than the run time for Algorithm \ref{algorithm:algo_WMMSECForm}, which is based on the closed form expressions of Theorem~\ref{theorem_h1}. Algorithm \ref{algorithm:algo_MMSECVX} has 2 orders of magnitude larger complexity than Algorithm \ref{algorithm:algo_WMMSECForm} either with or without power optimization. As they achieve similar MMF rates, we conclude that Algorithm \ref{algorithm:algo_WMMSECForm} is more advantageous than Algorithm \ref{algorithm:algo_MMSECVX}. In conclusion, 1-layer RS has the best MMF rate performance, Algorithm \ref{algorithm:algo_WMMSECForm} has the least complexity, and Algorithm \ref{algorithm:algo_SCA} provides a good tradeoff between complexity and MMF rates. It performs almost the same as 1-layer RS in MMF rates, and its complexity is only an order of magnitude larger than that of Algorithm \ref{algorithm:algo_WMMSECForm}. 

Note that, while solving \eqref{MMF_RS_all} with the MMSE approach, one could apply the KKT optimality conditions and the ordinary penalty method, instead of calling for CVX. However, the common and private rate expressions for rate splitting are complex and numerous, and finding the expressions the optimal precoders, receivers, weights and Lagrange multipliers as in Theorem \ref{theorem_h1} is complicated, keeping the complexity for 1-layer RS high.

\begin{figure}[t]
\vspace{-0.0in}
\centering
\hspace*{-0.0in}
\includegraphics[width=3.7in]{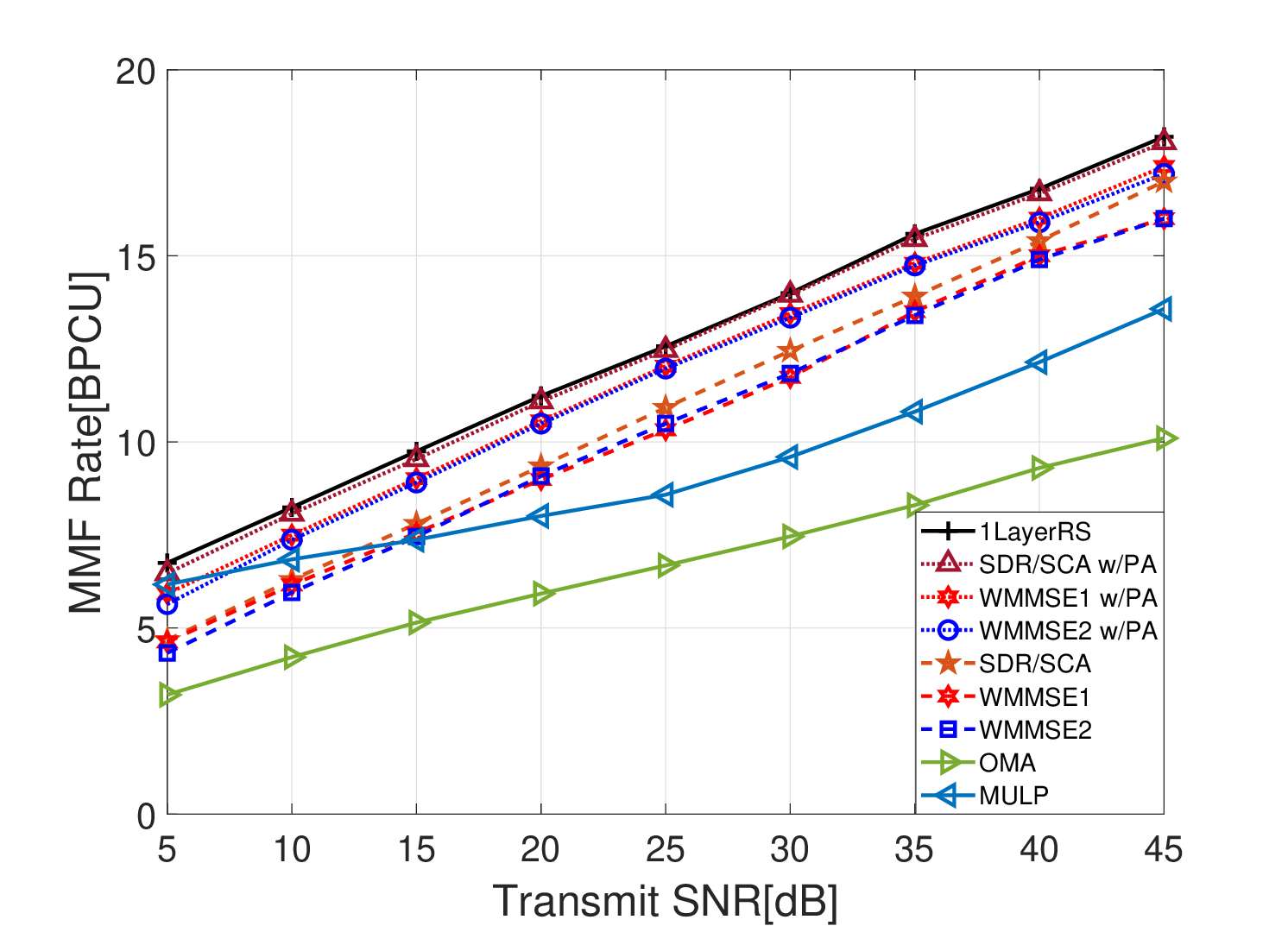}
\vspace{-0.1in}
\caption{MMF rates for different precoder schemes for $M=6, K=3, L=2$, $R_{k,l}^{th} = 0.1$ bpcu.}\label{fig:MMF_632}
\vspace{-0.00in}
\end{figure}

Figs. \ref{fig:MMF_332} and \ref{fig:MMF_442} are for overloaded systems. Fig. \ref{fig:MMF_632} shows how the MMF rates change, when $M$ is at least as large as $KL$. In the figure $M=6$, $K=3$ and $L = 2$ and $R_{k,l}^{th}= 0.1$ bpcu $\forall k \in \mathcal{K}$ and $\forall l \in \bar{\mathcal{L}}$. For this setting, the system is not overloaded, intense interference mitigation is not necessary, and benefits of rate splitting is less. Thus, SDR/SCA and WMMSE based schemes with or without power optimization are closer to 1-layer RS. RS and all the algorithms have full DoF equal to 1. OMA, by definition, still suffers from time division and its DoF is limited with 0.5. As the number of base station antennas is sufficient to serve all the users simultaneously, MULP also presents full DoF. This result is expected because the DoF for the designated beamforming scheme in \cite{Joudeh2017_RateSplittingforMaxMinFair} is proved to be 1, when there is a single user in each group and the number of base station antennas is  equal to the number of groups. However, MULP does not achieve this performance easily, its DoF result does not converge until 30 dB or higher. MULP is quite inefficient in interference mitigation and the additional threshold rate constraints for the weakest users in each group makes the MULP problem in \eqref{MMF_MULP_all} harder to solve especially for low to medium SNR.

\begin{figure}[t]
\vspace{-0.0in}
\centering
\hspace*{-0.0in}
\includegraphics[width=3.7in]{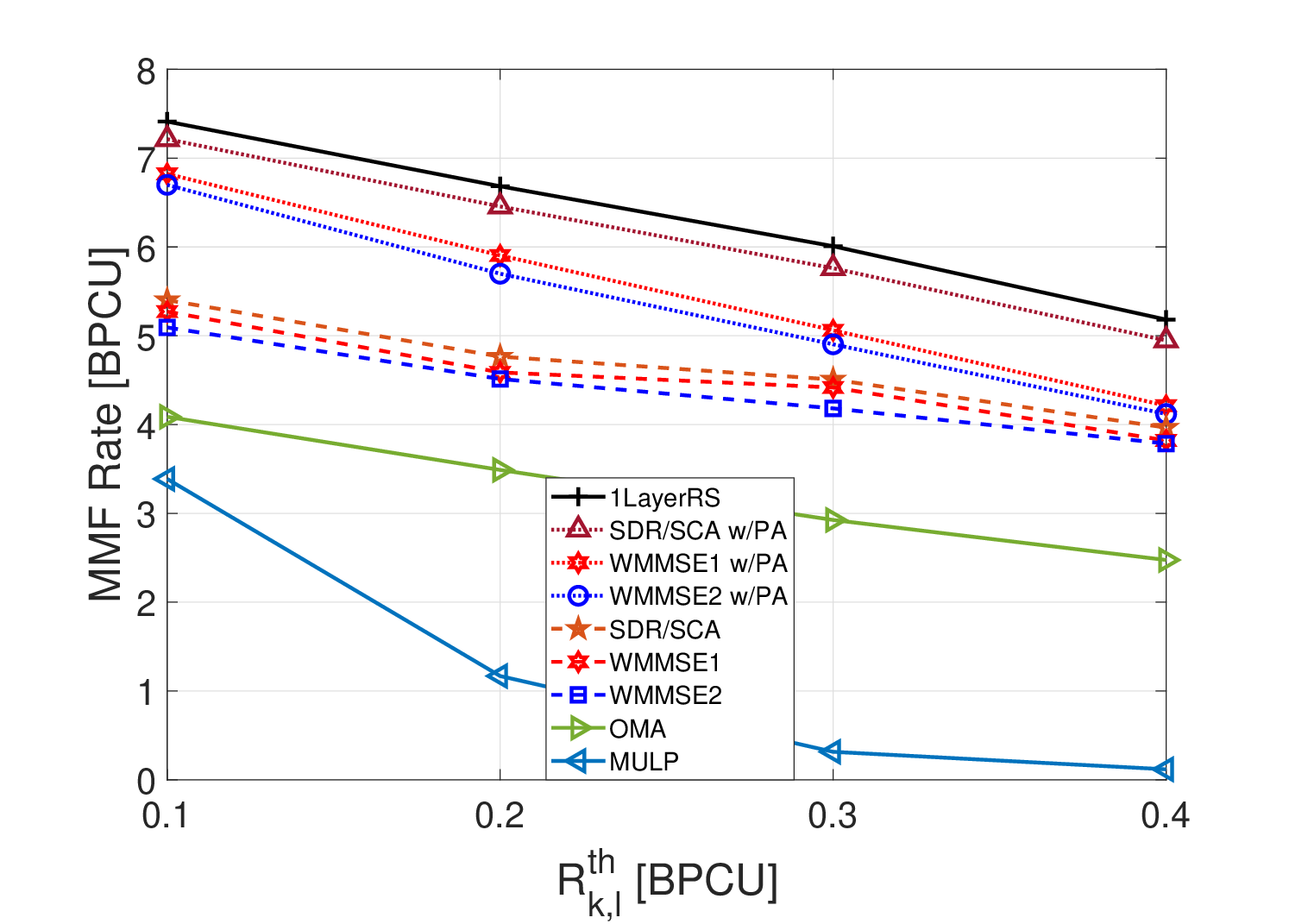}
\vspace{-0.1in}
\caption{MMF rates vs. $R_{k,l}^{th}$ for different precoder schemes for $M=K=3, L=2$. Transmit SNR is set to 15 dB.}\label{fig:MMF_332_diffRklth}
\vspace{-0.00in}
\end{figure}

\begin{figure}[t]
\vspace{-0.0in}
\centering
\hspace*{-0.0in}
\includegraphics[width=3.7in]{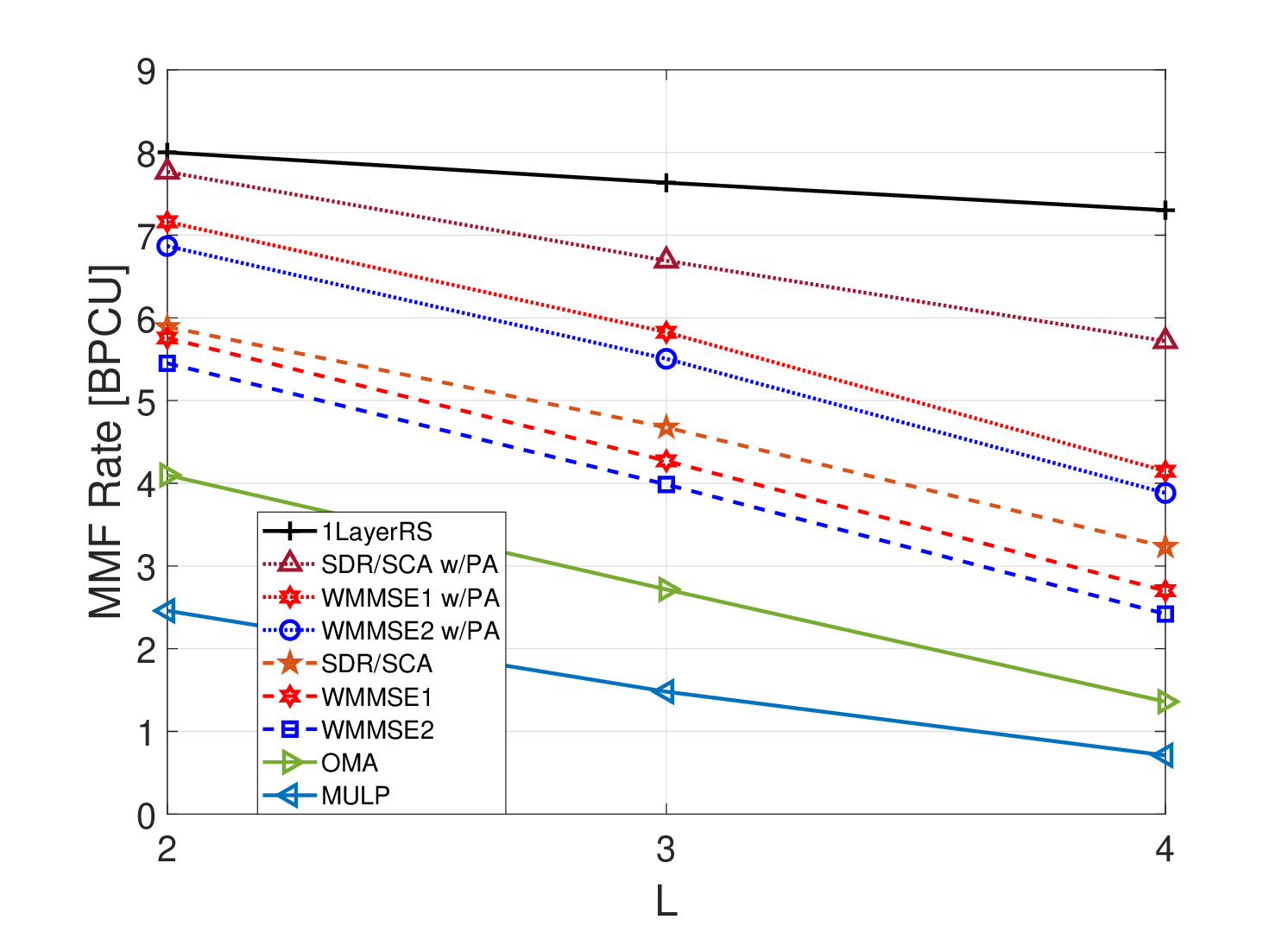}
\vspace{-0.1in}
\caption{MMF rates for different precoder schemes for $M=2, K=2, L=\{2,3,4\}$, $R_{k,l}^{th} = 0.1$ bpcu.}\label{fig:MMF_22_23}
\vspace{-0.00in}
\end{figure}

Fig. \ref{fig:MMF_332_diffRklth} shows that MMF rates decrease, when the threshold rates $R_{k,l}^{th}$, which is assumed to be the same $\forall k \in \mathcal{K}$ and $\forall l \in \bar{\mathcal{L}}$, increase from 0.1 to 0.4 bpcu for $M=3$, $K=3$ and $L=2$. The transmit SNR is set to 15 dB. Note that, one could expect MMF rate for OMA to be constant with increasing SNR. Time is divided into slots and all the strongest users are served in the last time slot, seemingly unaffected from all the other users. However, unless all the threshold rate constraints are satisfied, OMA is infeasible and MMF rate for OMA is zero. Therefore, MMF rate for OMA also decreases with increasing $R_{k,l}^{th}$. MULP rates decrease much faster than other schemes as the feasible set quickly shrinks with increasing $R_{k,l}^{th}$.

Fig. \ref{fig:MMF_22_23} presents the effect of increasing number of users in each cluster for $M=2, K=2, L=\{2,3,4\}$, $R_{k,l}^{th} = 0.1$ bpcu $\forall k \in \mathcal{K}$ and $\forall l \in \bar{\mathcal{L}}$. The decrease in MMF rates for 1-layer RS is much slower than all the other schemes as it provides excellent interference mitigation. 

\begin{figure}[t]
\vspace{-0.0in}
\centering
\hspace*{-0.0in}
\includegraphics[width=3.7in]{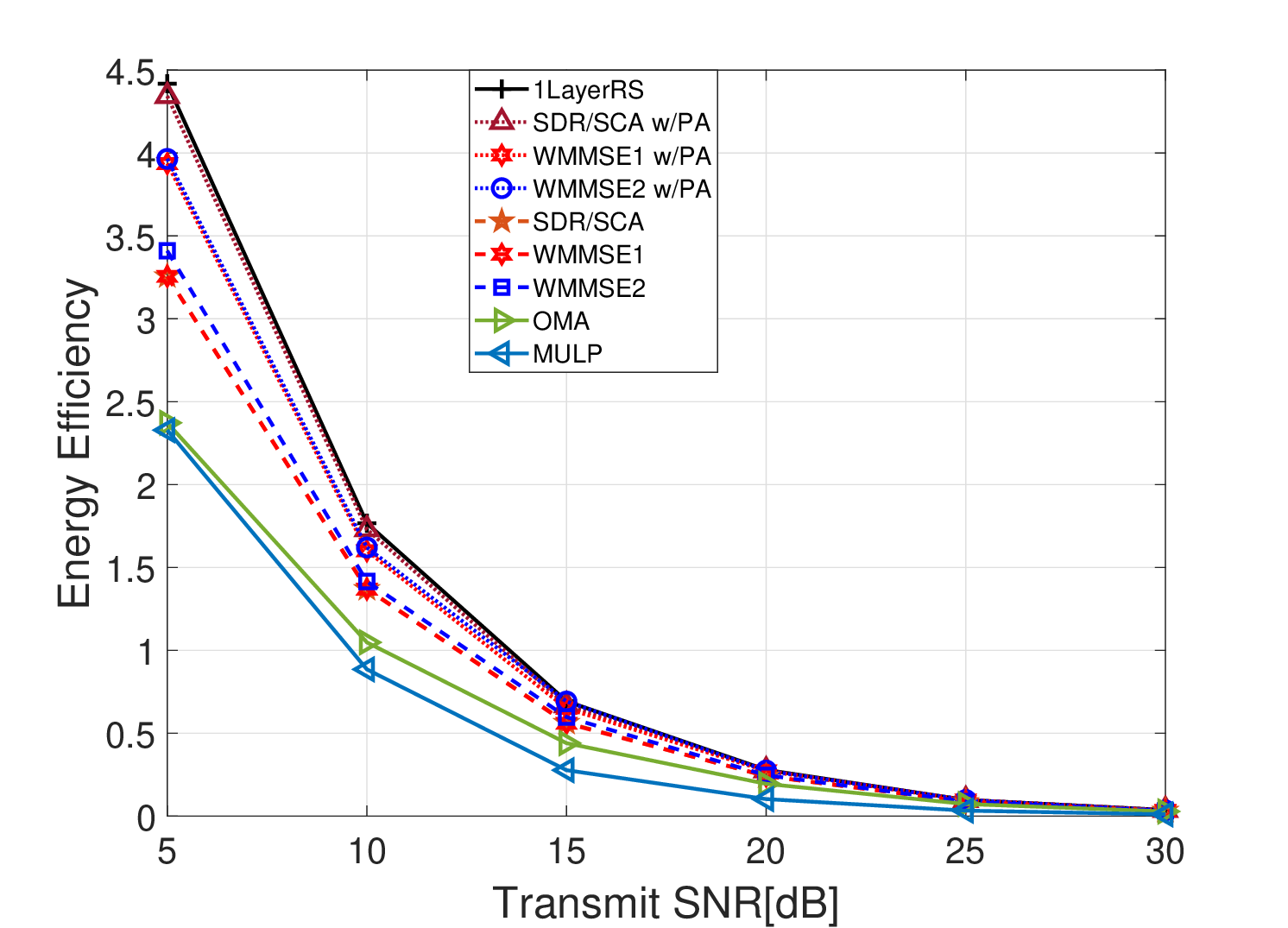}
\vspace{-0.1in}
\caption{Energy efficiency for different precoder schemes for $M=K=3, L=2$, $R_{k,l}^{th} = 0.1$ bpcu.}\label{fig:EE_332}
\vspace{-0.00in}
\end{figure}

Finally, in Fig.~\ref{fig:EE_332}, we compare all the schemes in terms of energy efficiency. Energy efficiency is defined as 
\begin{align}
EE &= \frac{\sum_{k\in\mathcal{K}}\sum_{l\in\mathcal{L}}R_{k,l} }{\Tr\left(\mathbf{P}\mathbf{P}^{H} \right)}
\end{align}for all precoding schemes. We observe that Algorithm \ref{algorithm:algo_SCA} has the same energy efficiency as 1-layer RS. The results show that the gap between the algorithms are smaller. Together with the results in Figs. \ref{fig:MMF_332} and \ref{fig:MMF_442}, and Table \ref{table_1}, we conclude that SDR/SCA with power allocation is an excellent scheme with high MMF rates, low complexity and high energy efficiency.

\section{Conclusion}\label{sec:conc}

We consider a joint precoder and power allocation design problem in downlink MIMO-NOMA to achieve max-min fairness among the strongest users in each cluster, while satisfying threshold rate constraints for all the other users. We propose 3 algorithms: (i) SDR/SCA, (ii) WMMSE1 and (iii) WMMSE2. The first algorithm is based on semi-definite relaxation and successive convex approximation, and the latter two are based on the relation between rate and minimum mean square error. WMMSE2 incorporates further simplifications in WMMSE1 based on the KKT optimality conditions and the ordinary penalty method. We compare our results with RS, OMA and MULP schemes. The results reveal that SDR/SCA scheme offers high MMF rates and superior energy efficiency at very low complexity. Future work includes designing precoders for imperfect channel state information and for finite block length channel coding. 

\begin{appendices}

 \section{}\label{derive_Pk}
 In this appendix, we prove Theorem \ref{theorem_h1}. Taking the derivative of the objective function $h$ in (\ref{lagrangian_h}) with respect to ${V}_{k, i \rightarrow l}$, then equating it to zero, we obtain 
 \begin{align}
 \alpha_{k,l} \mathbf{p}_k^H \mathbf{h}_{k,i}
&=  \sum_{j=l}^{L} \alpha_{k,j} \mathbf{h}_{k,i}^{H} \mathbf{p}_k \mathbf{p}_k^H \mathbf{h}_{k,i} {V}_{k, i \rightarrow l}    \nonumber \\
 &\:\quad +  \sum_{t=1, t\neq k}^{K} \mathbf{h}_{k,i}^{H} \mathbf{p}_t \mathbf{p}_t^H \mathbf{h}_{k,i} {V}_{k, i \rightarrow l} +  {V}_{k, i \rightarrow l}. \label{V_power}
 \end{align}Then, when $\eta_{k, i \rightarrow l} > 0$,
 \begin{align}
 {V}_{k, i \rightarrow l} &= \alpha_{k,l} \mathbf{p}_k^H \mathbf{h}_{k,i} T_{k, i \rightarrow l}^{-1}\label{V_rec_append}.
 \end{align}
 Secondly, taking the gradient of (\ref{lagrangian_h}) with respect to ${\mathbf{p}_k^H}$, and equating it to zero, we have the following equation
 \begin{align}
&\sum_{l=1}^{L} \sum_{i=l}^{L}\eta_{k, i \rightarrow l} b_{k, i \rightarrow l}  \mathbf{h}_{k,i} {V}_{k, i \rightarrow l}^{\ast}\alpha_{k,l}   \nonumber \\
&= \sum_{l=1}^{L} \sum_{i=l}^{L} \sum_{j=l}^{L} \alpha_{k,j} \eta_{k, i \rightarrow l} b_{k, i \rightarrow l}  \mathbf{h}_{k,i} {V}_{k, i \rightarrow l}^{\ast} {V}_{k, i \rightarrow l}\mathbf{h}_{k,i}^{H}\mathbf{p}_k \nonumber \\
&\quad +\sum_{\substack{t=1\\t\neq k}}^{K} \sum_{l=1}^{L} \sum_{i=l}^{L}\eta_{t,i \rightarrow l} b_{t,i \rightarrow l}
\mathbf{h}_{t,i} |{V}_{t,i \rightarrow l}|^2 \mathbf{h}_{t,i}^{H} \mathbf{p}_k + \beta \mathbf{p}_k.\label{Pk_power}
 \end{align}
 Then,
 \begin{align}
 \mathbf{p}_k &= \bigg[ \beta\mathbf{I} + \sum_{l=1}^{L} \sum_{i=l}^{L}\sum_{j=l}^{L} \alpha_{k,j} \eta_{k, i \rightarrow l} b_{k, i \rightarrow l}  \mathbf{h}_{k,i} |{V}_{k, i \rightarrow l}|^2 \mathbf{h}_{k,i}^{H} \nonumber \\
&\quad \:+ \sum_{t=1,t\neq k}^{K} \sum_{l=1}^{L} \sum_{i=l}^{L}\eta_{t,i \rightarrow l} b_{t,i \rightarrow l}
\mathbf{h}_{t,i} |{V}_{t,i \rightarrow l}|^2  \mathbf{h}_{t,i}^{H} \bigg]^{-1}\nonumber\\
&\:\quad \times \left[ \sum_{l=1}^{L} \sum_{i=l}^{L}\alpha_{k,l} \eta_{k, i \rightarrow l} b_{k, i \rightarrow l}  \mathbf{h}_{k,i} {V}_{k, i \rightarrow l}^{\ast}  \right].\label{Pk_append}
 \end{align}
 
To calculate $\beta$, we post-multiply both sides of (\ref{V_power}) by ${V}_{k, i \rightarrow l}^{\ast}\eta_{k, i \rightarrow l} b_{k, i \rightarrow l}$ and perform \\*$ \sum_{k=1}^{K} \sum_{l=1}^{L}\sum_{i=l}^{L}$ on both sides. We also pre-multiply (\ref{Pk_power}) with $\mathbf{p}_k^H$ and sum over $k, k=\{1,2,\ldots,K\}$. After calculating the trace of these two resulting equations, we observe that the left sides of both equations are equal. Then, the right sides are also equal to each other. As we assume that the power constraint in (\ref{pow_const}) is satisfied with equality we can find that 
 \begin{align}
 \beta &= \frac{1}{E_{tx}}\left[\sum_{k=1}^{K}\sum_{l=1}^{L}\sum_{i=l}^{L}\eta_{k, i \rightarrow l} b_{k, i \rightarrow l} |{V}_{k, i \rightarrow l}|^2 \right]. \label{beta_append}
 \end{align}

 \end{appendices}


\ifCLASSOPTIONcaptionsoff
  \newpage
\fi



%

\bibliographystyle{IEEEtran} 
\bibliography{journal.bbl}
\end{document}